\documentclass{aa}  

\usepackage{ulem}
\usepackage{graphicx}
\usepackage{url}
\usepackage{color}
\usepackage{xcolor}
\definecolor{xlinkcolor}{cmyk}{1,1,0,0}
\usepackage{threeparttable}
\usepackage[whole]{bxcjkjatype}
\usepackage{enumerate}
\usepackage{physics}
\usepackage{bm}
\usepackage{threeparttable}
\usepackage{txfonts}
\newcommand{\fullfigref}[1]{Figure \ref{#1}} 
\newcommand{\figref}[1]{figure \ref{#1}} 
\newcommand{\tabref}[1]{table \ref{#1}} 
\newcommand{\secref}[1]{section \ref{#1}}

\newcommand{\rmf}[1]{{_{\rm #1}}}
\begin{document} 
   \title{Simulations of two-temperature jets in galaxy clusters}

   \subtitle{I. Effect of jet magnetization on dynamics and electron heating}

   \author{T. Ohmura\inst{1,2,3}
          \and
          M. Machida\inst{3}
          }

   \institute{Institute for Cosmic Ray Research, The University of Tokyo, 5-1-5 Kashiwanoha, Kashiwa, Chiba 277-8582 Japan\\
              \email{tohmura@icrr.u-tokyo.ac.jp}
         \and
         Department of Physics, Faculty of Sciences, Kyushu University, 744 Motooka, Nishi-ku, Fukuoka 819-0395, Japan\\
         \and
        Division of Science, National Astronomical Observatory of Japan, 2-21-1 Osawa, Mitaka, Tokyo 181-8588, Japan\\
             \email{mami.machida@nao.ac.jp}
             }

   \date{accepted September 11, 2023}

 
  \abstract
   {
   Non-radiating protons in the radio lobes have an essential role to form the jet morphology which is shown by recent radio and X-ray observations.
   However, since protons and electrons are not always in energy equilibrium due to weak Coulomb coupling, it is difficult to estimate the energy contribution of protons for inflation of radio lobes. 
   }
   {
   The main focus of this study is to examine the effect of the variable model for electron heating by turbulence and shock waves on the thermal energy distribution of electron and proton.
   }
   {
   We performed two-temperature three-dimensional magnetohydrodynamic simulations of sub-relativistic jets in the galaxy cluster while varying jet magnetization parameters.
   Because the energy partition rate between electrons and protons in shock and turbulence is determined by plasma kinetic scale physics, our global simulations include electron instantaneous heating sub-grid models for shock waves and turbulence.
   }
   {
   We find that most of the bulk kinetic energy of the jet is converted into thermal energy of protons through both shocks and turbulence. Thus, protons are energetically dominant.
Meanwhile, thermal electrons stored in the lobe evolve toward energy equipartition with magnetic energy through turbulent dissipation.
We further estimated the radio power and the mechanical jet power of radio lobes following the same method as used for radio and X-ray observations, and compared these powers with that of the observed radio jets. 
The two-temperature model quantitatively explains the radiatively inefficient radio cavities, but cannot reproduce the radiatively efficient cavity, even for strong magnetized jets.
This implies that a significant population of pair-plasma is needed in the radiatively efficient radio cavities.
   }
   {}

   \maketitle
%
\section{Introduction}
Jets driven by the active galactic nuclei play a significant role in the galaxy and cluster evolution.
They propagate beyond the spatial scale of its host galaxy, and transport its own kinetic energy into the surrounding intracluster medium (ICM). 
This heating energy prevents ICM from catastrophic cooling and falling into the host galaxy. \citep{1994ARA&A..32..277F,2007ARA&A..45..117M,2012ARA&A..50..455F}.
To better understand and discuss the heating energy of ICM in a quantitative manner, estimating the total kinetic energy of the jet is of high importance.
Relativistic electrons and magnetic fields contained in the jet-driven lobes can be constrained by radio observations using the equipartition energy condition for the energy of cosmic ray and magnetic fields \citep{1984ApJ...285L..35P,2005AN....326..414B}.
It is, however, difficult to estimate the total kinetic power of the jet, as most of the radiation stems from non-thermal electron origin \citep{2020NewAR..8801539H}, i.e., the energy of thermal electrons and protons cannot arise directly from the source signals.

In the context of jets in galaxy clusters, the jet kinetic energy is constrained by observation of the X-ray cavity, which is produced as a result of the radio lobe inflation \citep{2000MNRAS.318L..65F, 2001ApJ...562L.149M}.
The minimum energy needed to form a cavity has been estimated as $E = P_{\rm cav} t_{\rm age} = 4pV$, where $P_{\rm cav}$, $t_{\rm age}$, $p$, and $V$ are the cavity power, source age, pressure of ICM observed by thermal X-ray, and volume of cavity, respectively.
\citet{2008ApJ...686..859B, 2004ApJ...607..800B} found that the mechanical power estimated from the X-ray cavity seem to be correlated with the radio luminosity (sum of core and lobes), and that cavity increases with radio luminosity, $P\rmf{cav} \propto P^{\alpha}\rmf{radio}$, where $0.35 \leq \alpha \leq 0.70$.
The median ratio of the mechanical power to radio luminosity (radiative efficiency) is $P\rmf{cav}/P\rmf{Radio} \sim 100$.
These results imply that the energy contribution of non-radiating protons is needed for cavity formation.
However, there is a large scatter in this relation.
For example, Cygnus A is the most radiatively efficient system, $P\rmf{cav}/P\rmf{Radio} \sim 1$.
Although there are several physical factors of these scatters, such as electron cooling, estimation of ages, plasma composition, and variable activity of AGNs, their contribution in creating this scatter is not understood.

Fluid simulations can follow highly complex jet flows and are a useful approach to examine the energy transport in the jet-ICM system. 
To approach the realistic conditions of radio lobes, some numerical models implement key physics, such as magnetic fields \citep{2019A&A...621A.132M}, special relativity \citep{1999ApJ...523L.125A}, difference plasma compositions \citep{2002MNRAS.331..615S,2014MNRAS.445.1462P}, cosmic ray electrons \citep{1999ApJ...512..105J, 2012ApJ...750..166M, 2021MNRAS.505.2267M}, and cosmic ray protons \citep{2010ApJ...725.1440M, 2017MNRAS.470.4530W}.
Several studies pointed out that magnetohydrodynamics (MHD) instabilities develop non-axisymmetric modes, namely Kelvin-Helmholtz modes \citep{1994A&A...283..655B}, Rayleigh-Taylor modes \citep{2013ApJ...772L...1M}, and current-driven kink modes \citep{2009ApJ...700..684M,2010MNRAS.402....7M,2015MNRAS.452.1089P}.
Thus, the non-linear evolution of these instabilities plays a significant role in the jet dynamics and their large-scale morphology \citep[e.g.,][]{2016MNRAS.461L..46T}.
In particular, the development of instabilities that cause the jet deceleration and/or jet disruption, could directly link to the physical reasons of the Fanaroff-Riley (FR) distinction \citep{1974MNRAS.167P..31F}.

Our previous studies \citep{2019Galax...7...14O,2020MNRAS.493.5761O} focused on the two-temperature plasma, where the electrons and protons in jets are not in thermal equilibrium, because Coulomb coupling is inefficient in the tenuous plasma \citep{1965RvPP....1..205B,1983MNRAS.204.1269S}.
We discuss the results of axisymmetric simulations with a constant fraction model for electron and proton heating.
Because thermal protons heated up at the internal shocks, and as the Coulomb coupling did not work effectively in the jets, the proton temperature was several times higher than electron temperature. 
Thus, we found that thermal protons support the expansion of the cocoon, rather than thermal electrons.

Recent theoretical studies clarified the physical picture of electron heating in the collisionless turbulence \citep{2010MNRAS.409L.104H,2019PNAS..116..771K,2020arXiv200404922K}.
These provide the variable model that describes the partition of heating energy between protons and electrons for the dissipation at the plasma kinetic scale.
The heating fraction between electrons and protons in this model is not constant, and represents an increasing function of the proton plasma-$\beta$, $\beta\rmf{p} \equiv 8\pi nkT\rmf{p}/B^2$.
However, in our previous simulation, we did not focus on the electron heating in turbulent conditions.
Our previous results indicated that the magnetic fields accumulate, and the proton plasma-$\beta$ decreased in the cocoon.
Furthermore, the magnetic field can be amplified locally by non-axisymmetric motion.
Therefore, the turbulence is the dominant heating source for thermal electrons, compared with shock heating.

Jet magnetization parameters, namely jet Alfv\'en Mach number ${\mathcal M}\rmf{A}$ and plasma-$\beta$ $\beta\rmf{gas}$, are important for both electron heating and dynamical evolution.
However, none of the studies considered two-temperature plasma, and thus failed to explore the effect of different jet magnetization on the electron heating with the development of MHD instabilities.
The purpose of this study is to examine the effect of the variable model for electron heating under turbulence on the distribution of electron temperature while varying jet magnetization parameters.

In this study, we present the results of three-dimensional, two-temperature MHD simulations of semi-relativistic jets in galaxy clusters.
In \secref{sec:2}, basic equations and sub-grid models for electron heating are presented.
We describe the setup of our simulation in \secref{sec:3}, and the results are presented in \secref{sec:4}.
In \secref{sec:5}, we discuss the observational implications of this study.
A brief summary of key findings is provided in \secref{sec:6}.
The appendix describes detailed numerical methods for solving entropy equations and the shock-finding algorithm. 

\section{Numerical Method} \label{sec:2}
\subsection{Basic equations}
Methods have been developed to incorporate electron thermodynamics into single-fluid simulations self-consistently in the context of the general relativistic MHD simulations for hot accretion flow \citep{2015MNRAS.454.1848R, 2017MNRAS.466..705S}.
In this work, we extend the single-temperature MHD code CANS+ \citep{2019PASJ...71...83M} to a two-temperature framework as follow the method in \citet{2017MNRAS.466..705S}.
The total gas (summed electrons and protons) evolves by the MHD equations in conservation form:
\begin{equation}
  \label{eq:conserved}
  \frac{\partial{\bm{U}}}{\partial{t}} +\nabla \cdot \bm{F} = \bm{S},
\end{equation}
where $\bm{U}$, $\bm{F}$, and $S$ are the vector of conserved quantities, the vectors of flux, and the vector of source term, respectively.
The conserved quantities read
\begin{equation}
  \bm{U} = \left(
             \begin{array}{c}
                \rho        \\
                \rho \bm{v} \\
                \bm{B}      \\
                E
             \end{array}
           \right) ,
\end{equation}
where $\rho$, $\bm{v}$ and $\bm{B}$ are the mass density, the bulk velocity, and the magnetic field, respectively. 
We assume the gas to hydrogen, such that $n_{\rm e} \sim n_{\rm p}$, where $n_{\rm e,p}$ are respectively the number density of proton and electron. Then, $\rho = m_{\rm e}n_{\rm e} + m_{\rm p}n_{\rm p} \approx m_{\rm p}n$.
The total energy $E$ and total pressure $p_{\rm T}$ are respectively
\begin{equation}
  E = \frac{p\rmf{p}}{\gamma\rmf{p}-1}  + \frac{p\rmf{e}}{\gamma\rmf{e}-1} + \frac{\rho v^2}{2} + \frac{B^2}{2} = \frac{p\rmf{gas}}{\gamma\rmf{gas}(T\rmf{p},T\rmf{e})-1} + \frac{\rho v^2}{2} + \frac{B^2}{2},
\end{equation}
\begin{equation}
  p\rmf{T} = p\rmf{p}+p\rmf{e}+\frac{B^2}{2} = p\rmf{gas}+\frac{B^2}{2},
\end{equation}
where $p\rmf{gas} = p\rmf{p} + p\rmf{e}$ is the gas pressure, which sums the proton pressure $p\rmf{p}$ and electron pressure $p\rmf{e}$.
The fluxes are then 
\begin{equation}
  \bm{F} = \left(
             \begin{array}{c}
              \rho \bm{v}  \\
              \rho \bm{v} \bm{v} + p_{\rm T} \bm{I} - \frac{1}{4\pi}\bm{B}\bm{B}  \\
               \bm{v}\bm{B}-\bm{B}\bm{v} \\
               (E+p_{\rm T})\bm{v}-\frac{1}{4\pi}\bm{B}(\bm{v}\cdot \bm{B})
             \end{array}
           \right)  ,
\end{equation}
The source terms are 
\begin{equation}
  \bm{S} = \left(
             \begin{array}{c}
               0 \\
               0  \\
               0 \\
               -q\rmf{rad}
             \end{array}
           \right)  ,
\end{equation}
$q\rmf{\rm rad}$ is the radiative energy loss rate.
In this study, we assume the radiation process as bremsstrahlung emission.
Notably, an adiabatic index of gas $\gamma\rmf{gas}$ and the radiative energy loss rate by bremsstrahlung radiation $q\rmf{rad}$ are functions of the electron and proton temperature.
The primitive variables of the above system of equations are:
\begin{equation}
    \bm{V} = (\rho, \bm{v}, p_{\rm gas}, \bm{B})^T.
\end{equation}
In addition to solving MHD equations, we solve the entropy equations of the two species to obtain each temperature.
The entropy equations of electrons and protons can be expressed as:
\begin{eqnarray}
  T\rmf{e}\frac{d(n\rmf{e}s\rmf{e})}{dt} &=& f\rmf{e}q\rmf{heat}+q\rmf{ie}-q\rmf{rad}, \label{eq:entropy1}\\
  T\rmf{p}\frac{d(n\rmf{p}s\rmf{p})}{dt} &=& (1-f\rmf{e})q\rmf{heat}-q\rmf{ie}, \label{eq:entropy2},
\end{eqnarray}
where $q\rmf{ie}$, $f\rmf{e}$, and $q\rmf{heat}$ are the energy transfer ratio via Coulomb coupling, the fraction of electron heating, and the dissipation heating rate, respectively.
The detailed procedures of numerical integration are described in Appendix \ref{app:1}.

We address the trans-relativistic regime for electrons, and use the following approximate entropy formula derived by \citet{2017MNRAS.466..705S} in Appendix A:
\begin{equation}
  \label{eq:ent_temp}
  s\rmf{e} \approx k \ln{ \left[ \frac{\theta\rmf{e}^{3/2} (\theta\rmf{e}+\frac{2}{5})^{3/2} }{\rho\rmf{e}} \right] },
\end{equation}
where $\theta\rmf{e} \equiv kT_{\rm e}/m_{\rm e}c^2$ is the dimensionless temperature.
In this approximate formula, we can easily obtain the electron temperature for given $s_{\rm e}$ and $\rho_{\rm e}$ as
\begin{equation}
    \label{eq:ent_temp_invert}
    \theta_{\rm e} \approx \frac{1}{5} \qty[ \sqrt{1+25 \qty{ \rho_{\rm e} \exp \qty (s_{\rm e} k^{-1})^{2/3} } } -1 ].
\end{equation}
The adiabatic index for electrons is calculated as follows:
\begin{equation}
  \label{ap_eq:gme}
   \gamma\rmf{e}(\theta\rmf{e}) = \frac{10 + 20\theta\rmf{e} }{6 + 15\theta\rmf{e}}.
\end{equation}
In contrast, protons are non-relativistic in this simulation.
Thus, we use the non-relativistic entropy formula for protons:
\begin{equation}
  \label{eq:ent_temp_ion}
  s\rmf{p} = k \ln{ p\rmf{p} \rho\rmf{p}^{-\gamma\rmf{p}}},
\end{equation}
where $\gamma\rmf{p} = 5/3$.
The thermal energies of protons, electrons, and gas are as follows:
\begin{equation}
  \label{ap_eq:eos}
  u_{\rm p}= \frac{p_{\rm p} }{\gamma_{\rm p}-1},~~u_{\rm e}= \frac{p_{\rm e} }{\gamma_{\rm e}(T\rmf{e})-1},~~u_{\rm gas}= u\rmf{p} + u\rmf{e}= \frac{p_{\rm gas} }{\gamma_{\rm gas}(T\rmf{p},T\rmf{e})-1}.
\end{equation}
From the relationship between the gas pressure and gas thermal energy, the effective adiabatic index for the gas can be calculated as \citep{2015MNRAS.454.1848R}
\begin{equation}
  \gamma\rmf{gas}(T\rmf{p},T\rmf{e}) = 1 + (\gamma\rmf{e}-1)(\gamma\rmf{p}-1) \frac{1+T\rmf{p}/T\rmf{e}}{(\gamma\rmf{p}-1) + (\gamma\rmf{e}-1)T\rmf{p}/T\rmf{e}}.
  \label{eq:gamma_gas}
\end{equation}

\subsection{Sub-grid models of electron heating}
We consider two sub-grid models for the fraction of electron heating $f\rmf{e}$.
One model represents turbulence heating $f\rmf{e,turb}$, and another model represents the shock heating $f\rmf{e,shock}$.
Therefore, $f\rmf{e}$ is determined by plasma properties at each simulation grid.
First, we identify a shock zone by shock-finding method based on \citet{2003ApJ...593..599R} and \citet{2015MNRAS.446.3992S} (see detail in Appendix \ref{ap_sec:shock_find}).
The fraction of shock heating is adopted only in the shock zone, and the other region is adopted by the fraction of turbulence heating ,i.e.,
\begin{equation}
  f\rmf{e}(x,y,z) =  \begin{cases}
                    f\rmf{e,shock} & ({\rm for~shock~zone}) \\
                    f\rmf{e,turb}  & ({\rm for~otherwise})
                      \end{cases}
\end{equation}
Note that part of the dissipation energy is small in the region of laminar flow, and hence the heating fraction of turbulence heating spontaneously works in the turbulence zone.

For the shock zone, we model a constant electron heating fraction, $f\rmf{e,shock} = 0.05$.
This value is justified by the observation data in the solar system and supernova remnants shocks \citep{2015A&A...579A..13V}. 
Furthermore, some theoretical simulations, based on particle-in-cell (PIC) simulation, indicate that electrons irreversibly heat up, while protons are primarily heated during collisionless shocks \citep{2010PhPl...17d2901M,2018ApJ...858...95G,2019MNRAS.485.5105C,2020ApJ...900L..36T}.
The validity of this parameter is previously discussed in Section 4.1 of \citet{2020MNRAS.493.5761O}.

For electron to proton heating rates of MHD turbulence, there are two models proposed by \citep[H10][]{2010MNRAS.409L.104H} and \citep[K19][]{2019PNAS..116..771K}.
The model comparison between H10 and K19 is shown in Appendix \ref{sec:appendC}.
Both the models are based on the gyrokinetics approach to damping weakly collisional MHD turbulence.
H10 provided the heating model derived from the linear theory for the first time, while K19 treated the nonliner evolution of the turbulence by numerical simulations.
Thus, K19 would be favored. 
We use the K19 for the unshocked zone:
\begin{equation}
  \label{ch5_eq1:kawazura}
  \frac{Q\rmf{p}}{Q\rmf{e}} = \frac{35}{1+(\beta\rmf{p}/15)^{-1.4} {\rm e}^{-0.1 T\rmf{e}/T\rmf{i}}} + \frac{P\rmf{compr}}{P\rmf{AW}},
\end{equation}
where $P\rmf{compr}$ and $P\rmf{AW}$ are the compressive energy injection and Alfv\'{e}nic energy injection, respectively.
Therefore, the fraction of electron heating $f\rmf{e}$ is defined by:
\begin{equation}
  f\rmf{e,turb} = \frac{Q\rmf{e}}{Q\rmf{p}+Q\rmf{e}}.
\end{equation}
It is a difficult task to estimate the ratio of the compressive energy injection and the Alfv\'{e}nic energy injection in fluid simulations.
Therefore, we assume pure Alfv\'{e}nic turbulence (i.e., $P\rmf{compr}/P\rmf{AW} \to 0$).
Notably, this assumption leads to an overestimation of the amount of electron heating.
This heating model represents a weak dependence of the temperature ratio $T\rmf{e}/T\rmf{p}$, but a strong dependence of proton plasma beta $\beta\rmf{p}$.
In the case of $\beta\rmf{p} < 1$, most of the dissipation energy is absorbed by electrons, and vice versa.

\section{Simulation setup} \label{sec:3}
We carried out the two-temperature 3D MHD simulations in Cartesian coordinates with the z-axis pointing along the jet direction.
The computational domain is $x\in (-L_x/2,L_x/2)$, $y\in (-L_y/2,L_y/2)$, and $z\in (0,L_z)$, where $L_x$, $L_y$, and $L_z$ denote the length of the computational domain.
We use a uniform mesh of $(N_x,N_y,N_z)$ with size $\Delta_x = \Delta_y= \Delta_z = 0.1~{\rm kpc}$.
The grid number and length of the computational domain are given in \tabref{ch5_tab:model}.
We permit the backflow to escape from the boundary at $z = 0$.
Therefore, the absorbing boundary condition applies to the $xz-$ plane at $z=0$.
Other boundaries are imposed on the free-boundary condition.

\subsection{Initial condition}
To study the interaction between jets and the ICM, we initialize the surrounding ICM in the form of $\beta$ profile \citep{1962AJ.....67..471K}.
Our cluster model is roughly consistent with the environment of Cygnus A from {\it Chandra} X-ray data \citep{2006ApJ...644L...9W,2002ApJ...565..195S}.
The density profile of ICM is given by
\begin{equation}
  n(r) = \frac{n_0}{\left[ 1+(r/r\rmf{c})^2 \right]^{3\beta'/2}},
\end{equation}
where $r=\sqrt{x^2+y^2+z^2}$, $n_0$, $r\rmf{c}$ and $\beta'$ are the radius, core density, core radius, and ratio of the specific energy in galaxies to the specific thermal energy in the ICM, respectively.
We set $\beta' = 0.5$, $r\rmf{c} = 20$ kpc, and $n_0 =$ 0.05 ${\rm cm^{-3}}$.
We also assume that our atmosphere is initially isothermal with the temperature $kT\rmf{p} = kT\rmf{e} = 5$ keV.
We employ the uniform magnetic field $B\rmf{ICM}$ that is parallel to the z-axis, and $B\rmf{ICM} = 0.44~\mu{\rm G}$.
The blue lines of \figref{fig1} show the initial density (top panel) and pressure (bottom panel), respectively.

\subsection{Jet model}
Focusing on the effect of jet magnetization on electron heating and jet stability, we carried out simulations with different magnetic field strengths.
We modeled magnetized supersonic sub-relativistic flows to be consistent with the outburst energy of Cygnus A jets, $0.6-0.8 \times 10^{46}$ ${\rm erg~s^{-1}}$ \citep{2004ApJ...607..800B,2018ApJ...855...71S}.
The injected jet power can be written as follows:
\footnotesize
\begin{equation}
    L\rmf{jet} = L\rmf{kin} + L\rmf{th}+ L\rmf{mag} = \pi r\rmf{jet}^2 v\rmf{jet} \left( \frac{1}{2} m\rmf{p} n\rmf{jet} v^2\rmf{jet} + \frac{n\rmf{jet} kT\rmf{p,jet}}{\gamma\rmf{p}-1} + \frac{n\rmf{jet} kT\rmf{e,jet}}{\gamma\rmf{e}-1} + \frac{B^2\rmf{jet}}{8\pi} \right).
    \label{ch5_eq:jet_power}
\end{equation}
\normalsize
We set the kinetic power as $L\rmf{kin} = 5.5 \times 10^{45}$ erg ${\rm s^{-1}}$, and the thermal power as $L\rmf{th} = 4.4 \times 10^{44}$ ${\rm erg~s^{-1}}$.

\begin{figure}
   \begin{center}
     \includegraphics[width=1.0\columnwidth, bb= 0 0 460.8 345.6 ]{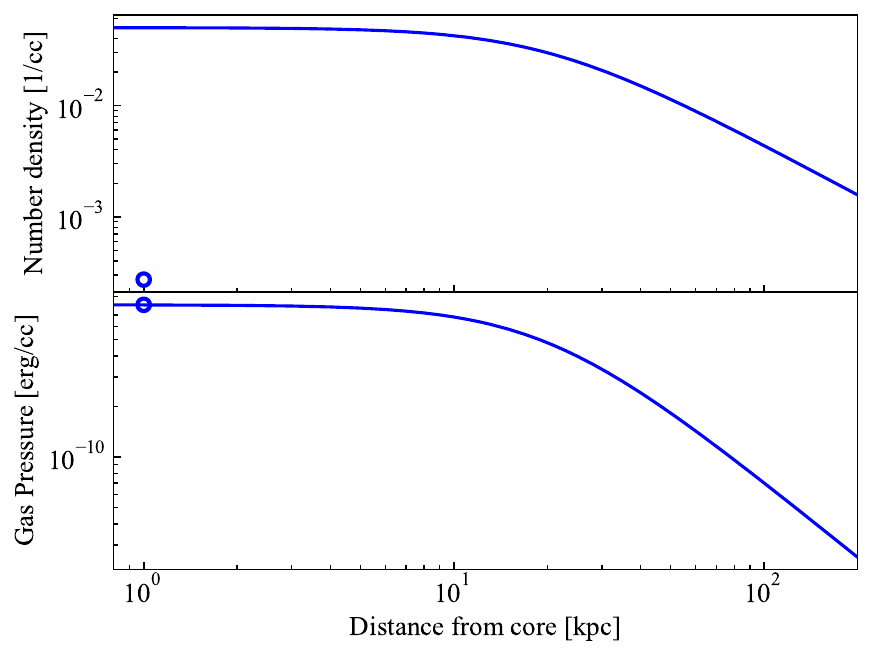}
     \caption{Number density ({\bf Top}) and gas pressure ({\bf Bottom}) profiles of initial ICM as a function of radius. Blue dots represent the jets number density and the jets gas pressure, respectively.  The initial gas temperature of ICM is 5 KeV over the entire simulation domain.}
     \label{fig1}
   \end{center}
 \end{figure}

To generate the jet beams, we injected supersonic and magnetized flows inside a constant cylindrical nozzle at the origin.
The radius and length of the nozzle are $1$ kpc and $1.2$ kpc, respectively.
Although it is difficult to determine the real flow radius of jets by observation. Radio observation indicates that the beam radius of Cyguns A is 0.1 to 1.0 kpc at 1 kpc from the central engine \citep{2019ApJ...878...61N}.
We assume that the jet temperature and the velocity are $T\rmf{p} = T\rmf{e} = 10^{10}$ K and $v\rmf{jet} = 0.3c$, respectively.
Thus, the internal sonic Mach number is ${\mathcal M} = 6.2$.
Our jet models satisfy the condition that the thermal pressure ratio between the jet and the ICM in the launching region is unity.
In \figref{fig1}, the blue dots for each panel denote the density and pressure in the jet injection region, respectively.
A small-amplitude (1 percent) random pressure perturbation for the injection flow is adapted to  model non-axisymmetric features.
We list common parameters for ICM and jets in \tabref{ch5_tab:common}.

The jets have a purely toroidal magnetic field $B_{\phi} = B\rmf{jet} \sin^4{(\pi r/r\rmf{jet})}$.
As shown in \tabref{ch5_tab:model}, Models A, B, and C have different values of gas plasma $\beta\rmf{gas}$ equal to 1, 5, 100, respectively.
In these case, the amplitudes of the injected magnetic field are $B_{\rm jet} = 138,~62,~14~\mu$G for models A, B, and C, respectively.

Our models are matter-dominated jets, i.e., the kinetic energy of jets exceeds the Poynting flux energy.

\begin{table*}
  \begin{center}
  \caption{Numerical Models}
  \label{ch5_tab:model}
    \begin{tabular}{c c c c c c}
      \hline
      Model &  $\beta\rmf{gas,jet}$ & ${\mathcal M_{\rm A}}$ & $B\rmf{jet}$ $[\mu$G] &$L_x \times L_y \times L_z$ [kpc] & $N_x \times N_y \times N_z$ \\ \hline\hline
      A &  1   & 4.9 & 138  & $64\times65\times96$ & $640\times650\times960$ \\
      B &  5   & 11  & 62   & $64\times64\times96$ & $640\times640\times960$ \\
      C &  100 & 49  & 14   & $64\times65\times96$ & $640\times650\times960$ \\ \hline
    \end{tabular}
\end{center}
\end{table*}

\begin{table}
  \begin{center}
  \caption{Jets and ICM common setup parameters}
  \label{ch5_tab:common}
    \begin{tabular}{c c c}
      \hline\hline
      Jet speed               & $v_{\rm jet}$     & 0.3$c$ \\
      Jet gas temperature     & $T_{\rm g,jet}$   & $10^{10}$ K \\
      Jet Kinetic energy      & $L_{\rm kin}$     & $5.0\times10^{45}$ erg ${\rm s^{-1}}$ \\
      Jet thermal energy      & $L_{\rm th}$      & $4.4\times10^{44}$ erg ${\rm s^{-1}}$ \\
      Jet radius              & $r_{\rm jet}$     & 1 kpc \\
      Jet Sonic Mach Number   & ${\mathcal M}_{\rm jet}$ & 6.2 \\ \hline
      ICM temperature         & $T_{\rm ICM}$     & 5 keV \\
      Core density            & $  n_{0}$         & $5\times10^{-2}~{\rm cm^{-3}}$ \\
      Core radius             & $r\rmf{c}$        & 20 kpc \\
      Core parameter          & $\beta'$           & 0.5 \\
      ICM magnetic field      & $B_{\rm z, ICM}$  & 0.44 $\mu G$     \\
      \hline\hline
  \end{tabular}
\end{center}
\end{table}

\section{Results} \label{sec:4}
We conducted simulations with various magnetic energies to investigate the effect of jet dynamics and electron heating.
First, we focus on the effect of the magnetic field strength on the jet dynamics.
The strength of the magnetic field affects the development of instability such as kink, Kelvin-Helmholtz, and Rayleigh-Taylor modes.
The electron heating model for turbulence is the function of plasma-$\beta\rmf{p}$, and hence magnetization strongly affects the electron temperature distributions.
Next, we report the time evolution for thermal electrons and protons in the jet lobe.

\subsection{Overall morphology and beam stability}
\begin{figure*}
   \begin{center}
     \includegraphics[width=1.0\textwidth]{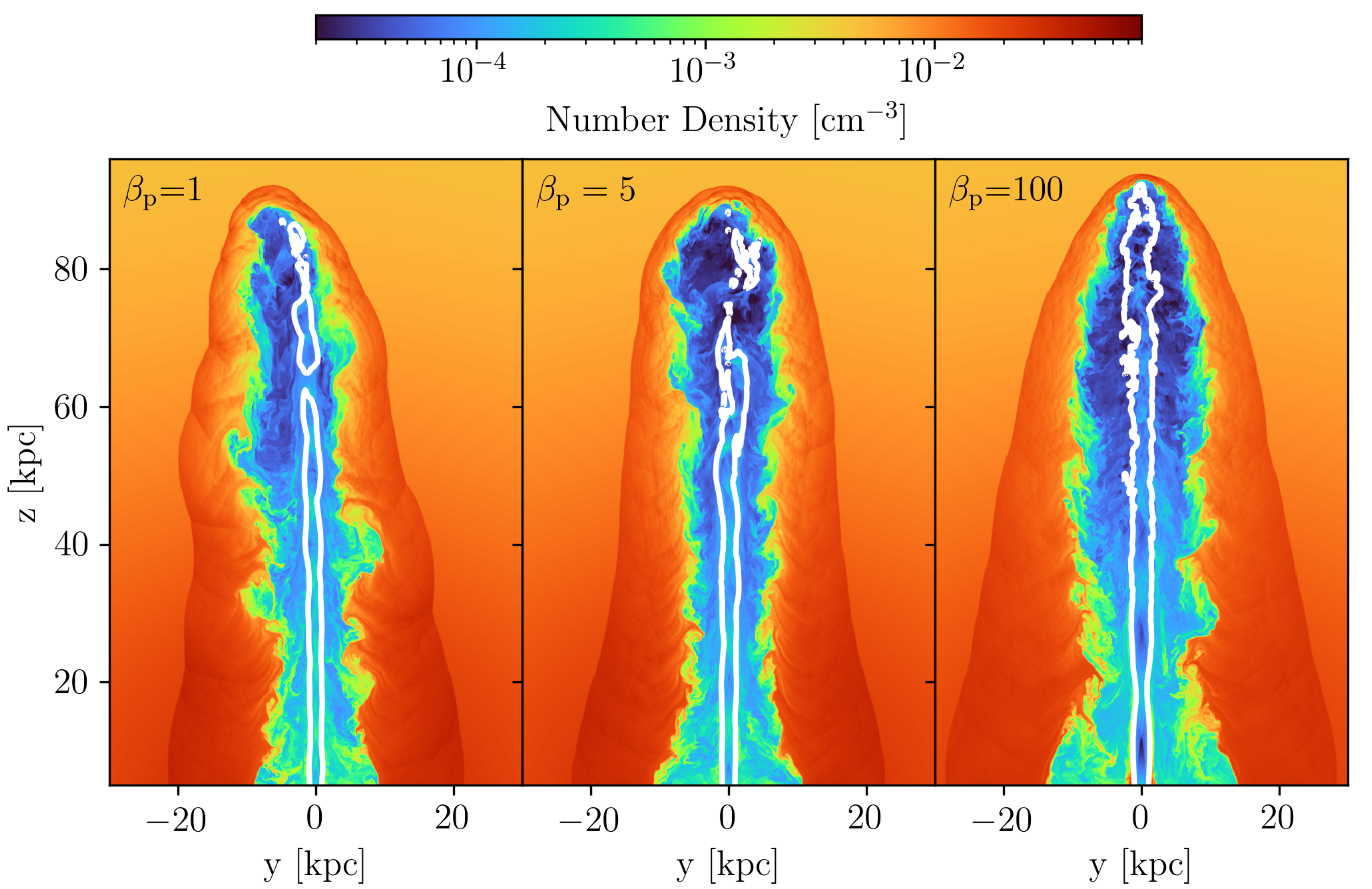}
     \caption{Slices (in the $y-z$ plane) of number density distribution for model A, B, and C at $t = 9.52$, $9.94,$ and 13.02 Myr, respectively. white lines represent contours of the z-component of velocity $v_{z} = 0.5v_{\rm jet}~(0.15c)$.}
     \label{fig2}
   \end{center}
 \end{figure*}

\fullfigref{fig2} shows the density slice in the $yz-$plane of $x = 0$ kpc at the end of the simulation ($t = 9.52,~9.94,$ and 13.02 Myr) for models A, B, and C.
The shocked-ICM, compressed by the bow shock, and the low-density cocoon are formed.
We observe that the number density of the cocoons is very low, $\sim 10^{-4}$ ${\rm cm^{-3}}$, such that Coulomb coupling is inefficient.
The jet beam reaches jet tip despite suffering MHD instabilities (see red contours in \figref{fig2}), and a terminate shock is formed at the end of the jet for all models.
We find that the non-axisymmetric mode is developed in model A and B, as the shapes of the bow shock are affected by the bending motion of the jets for both models.
Strong pressure waves are generated at the termination of the beam, which push up the bow shock.

For models A and B, the jet develops a pronounced helical shape that is characteristic for kink instabilities.
Although our jet has a purely toroidal magnetic field at the launching region, a helical component is generated during jet propagation.
The timescale of the development of an external kink mode corresponds to the Alfv\'{e}n crossing time in the beam \citep{2008A&A...492..621M,2009ApJ...700..684M},
\begin{equation}
  \tau_{\rm kink} \sim \frac{2\pi r_{\rm jet}}{v_{\rm A,\phi}},
\end{equation}
where $v_{A,\phi}$ is the azimuthal Alfv\'{e}n velocity.
A fluid element in the beam has roughly a constant velocity, $v_{\rm jet}$.
We verify that the bulk beam velocity changes slightly, but roughly maintains an injection velocity.
Therefore, the kink mode develops after the jets propagate to the distance $l\rmf{kink} \sim v\rmf{jet} \tau \rmf{kink}$.
For models A and B, the distances $l\rmf{kink} \sim 40,~70$ kpc are respectively within the simulation domain.
Meanwhile, $l\rmf{kink} \sim 300$ kpc is larger than the length of the simulation domain, and hence the model C jet is not expected to develop the kink mode in simulation time.

\begin{figure}
  \begin{center}
     \includegraphics[width=1.0\columnwidth]{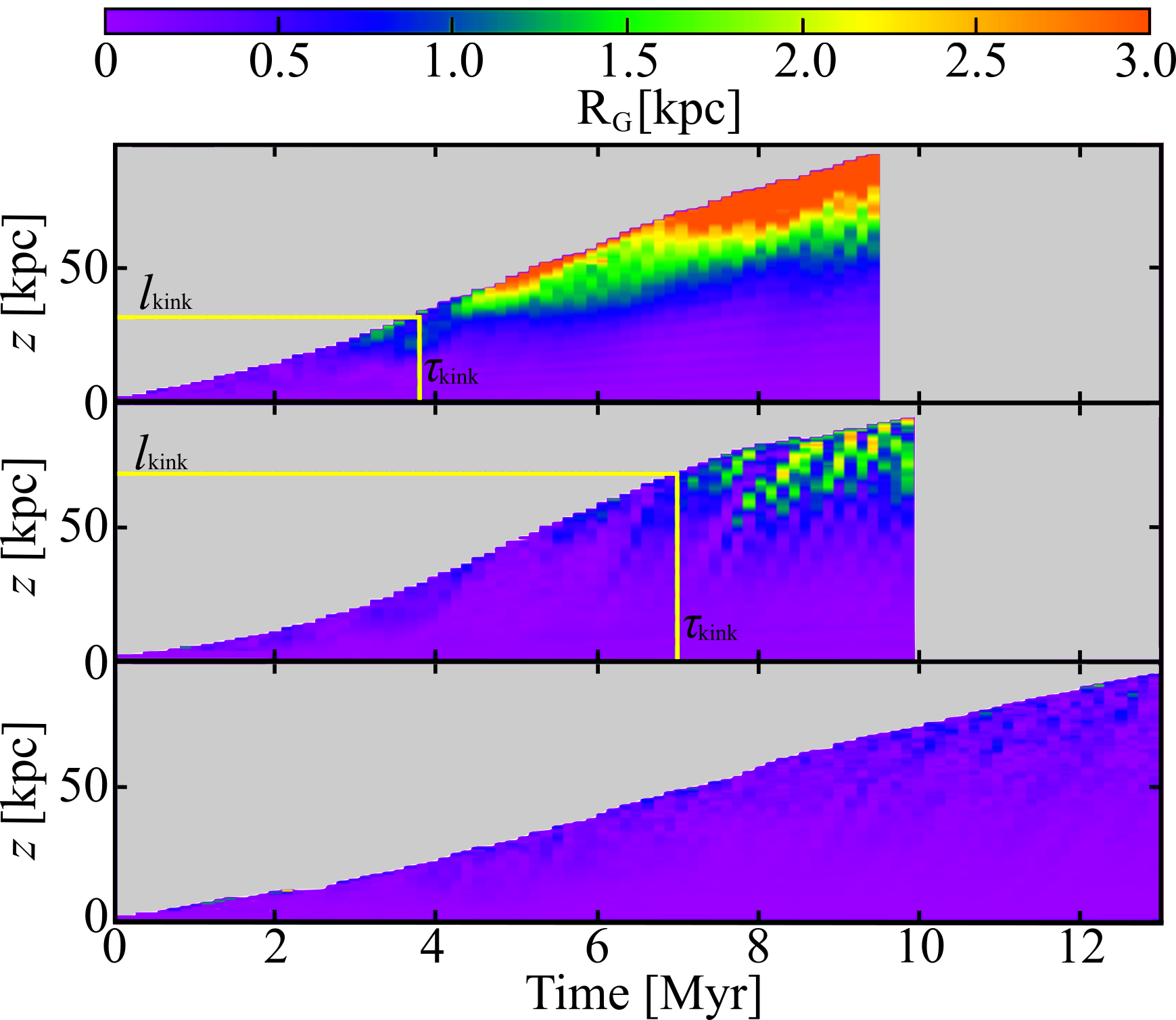}
     \caption{ Two-dimensional distribution maps of beam barycenter $R_{\rmf G}$ as a function of time for model A ({\bf top}), B ({\bf middle}), and C ({\bf bottom}), respectively. Horizontal yellow solid lines depict the distance $l\rmf{kink}$ for the development of the external kink model, and vertical yellow solid lines are the time, $\tau_{\rm kink}$, at which the jets propagate to distance $l\rmf{kink}$.  }
     \label{fig3}
  \end{center}
 \end{figure}

Figure \ref{fig3} shows the two-dimensional distribution maps of the beam barycenter $R_{\rmf G}$, which is described as the distance from the origin, as function of time for each model.
The large value of the barycenter indicates development of a non-axisymmetric mode.
We compute the beam barycenter as follows:
\begin{equation}
  R_{\rm G}(t,z) = \frac{\int_x \int_y r v\rmf{z}(x,y,z,t) dxdydz}{\int_x \int_y v\rmf{z}(x,y,z,t) dxdydz}\ \ \ {\rm for}\ \ \  v\rmf{z} > 0.8 v\rmf{jet}.
\end{equation}
Similar analysis is performed by \citet{2013MNRAS.436.1102M}.
We see that $l\rmf{kink}$ is a good indicator of the kink instability for both models A and B.
After the jet propagates to the distance $l\rmf{kink}$, the barycenter is larger than 3 kpc for both models.
For model B, the time at which the jet propagates to the distance $l\rmf{kink}$ is 7.5 Myr, which corresponds to the time when it starts to decelerate.
Thus, the non-axisymmetric mode developed by the kink instability induces deceleration by increasing the size of the jet head.
Meanwhile, because the jet does not suffer the kink mode, the barycenter for model C is within 1 kpc in simulation time, i.e., the jet propagates straight.

\begin{figure*}
   \begin{center}
     \includegraphics[width=1.0\textwidth]{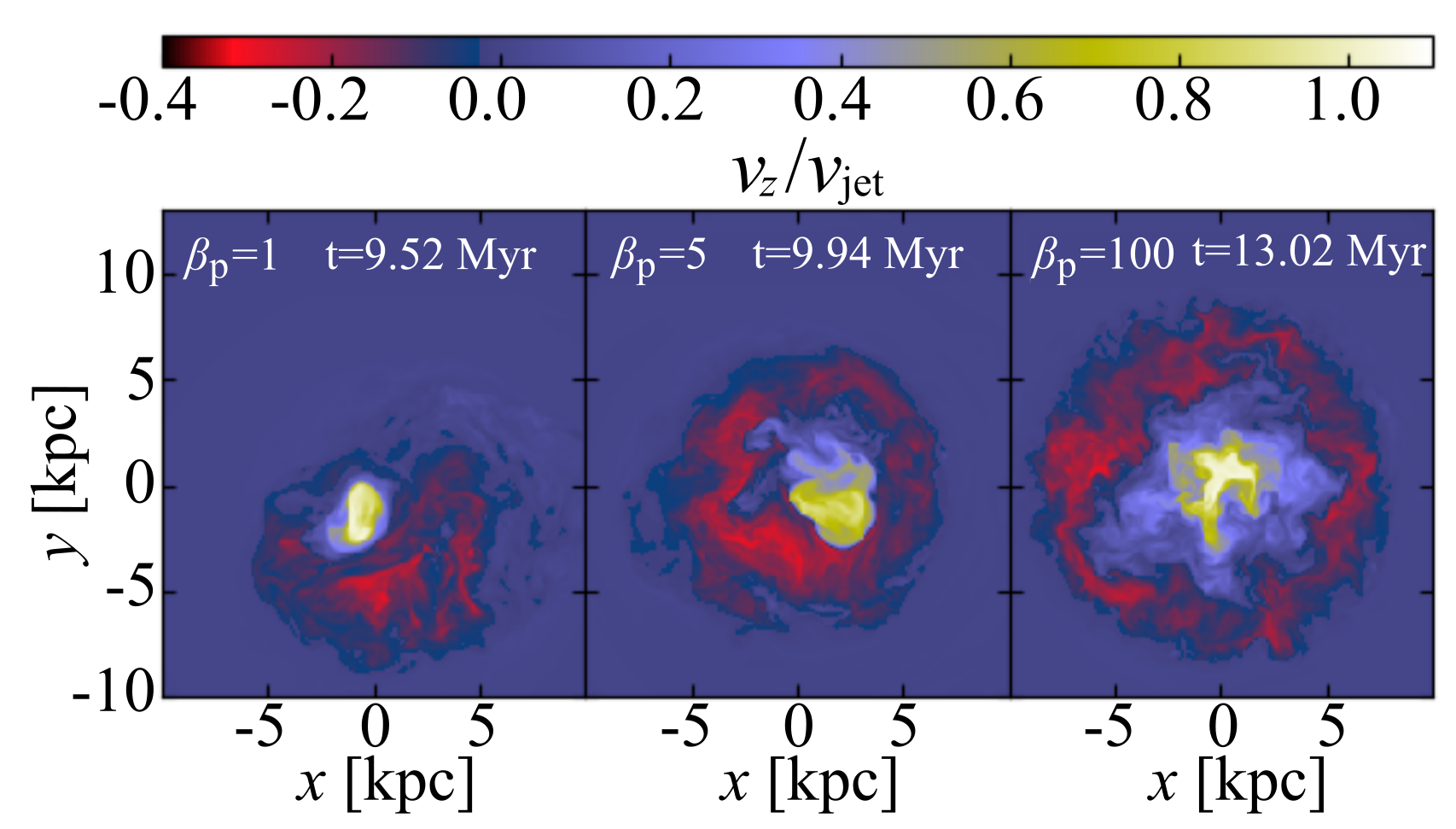}
     \caption{Slices (in the $x-y$ plane at $z = 80$ kpc) of the distribution of the z-direction component of velocity for models A, B, and C.}
     \label{fig4}
   \end{center}
 \end{figure*}

Magnetic fields also play an important role in the suppression of the Rayleigh-Taylor instability.
\fullfigref{fig4} shows the $v_z$ distribution on the xy-plane at $z=70$ kpc for model A, B, and C.
We observe that kink instability for models A and B bend the jet away from its initial launch axis ($x=y=0$).
Further, for models A and B, the jets clearly separate between the beam flow (yellow region) and cocoon gas (red region), i.e., the low mixing ratio between the beam and cocoon gas.
Meanwhile, the low-magnetized jet of model C is not in the development of kink instability, but in that of the Rayleigh-Taylor and Kelvin-Helmholtz instabilities.
Similar results have been presented for relativistic MHD jet simulations in \citet{2010MNRAS.402....7M} and \citet{2020MNRAS.499..681M}.
In addition to this, our results are supported by the linear stability analysis for a relativistic non-rotating jet, that indicates a strong magnetic field can suppress a growth rate of the Kelvin-Helmholtz instability \citep{2013MNRAS.434.3030B}. 
The onset condition of the Rayleigh-Taylor instability is given analytically by $\rho\rmf{jet} > \rho\rmf{cocoon}$ in the hydrodynamic case.
Notably, \citet{2019MNRAS.488.4061K} found analytically that jets are stable for the Rayleigh-Taylor instability mode when ${\mathcal M_{A}} < 40$.
Thus, the jet is in an unstable mode of the Rayleigh-Taylor instability for model C (see \figref{fig2} and \tabref{ch5_tab:model}).
The right panel of \figref{fig4} shows the Rayleigh-Taylor and the Kelvin-Helmholtz mode forming a large cross-section, $\sim 10$ kpc, of positive velocity field ($v_z >0$: from blue to yellow region), and finger-like structures.
Owing to the high-mixing ratio between the beam and cocoon gas, the jet of model C is decelerated.

\begin{figure*}
   \begin{center}
     \includegraphics[width=1.0\textwidth]{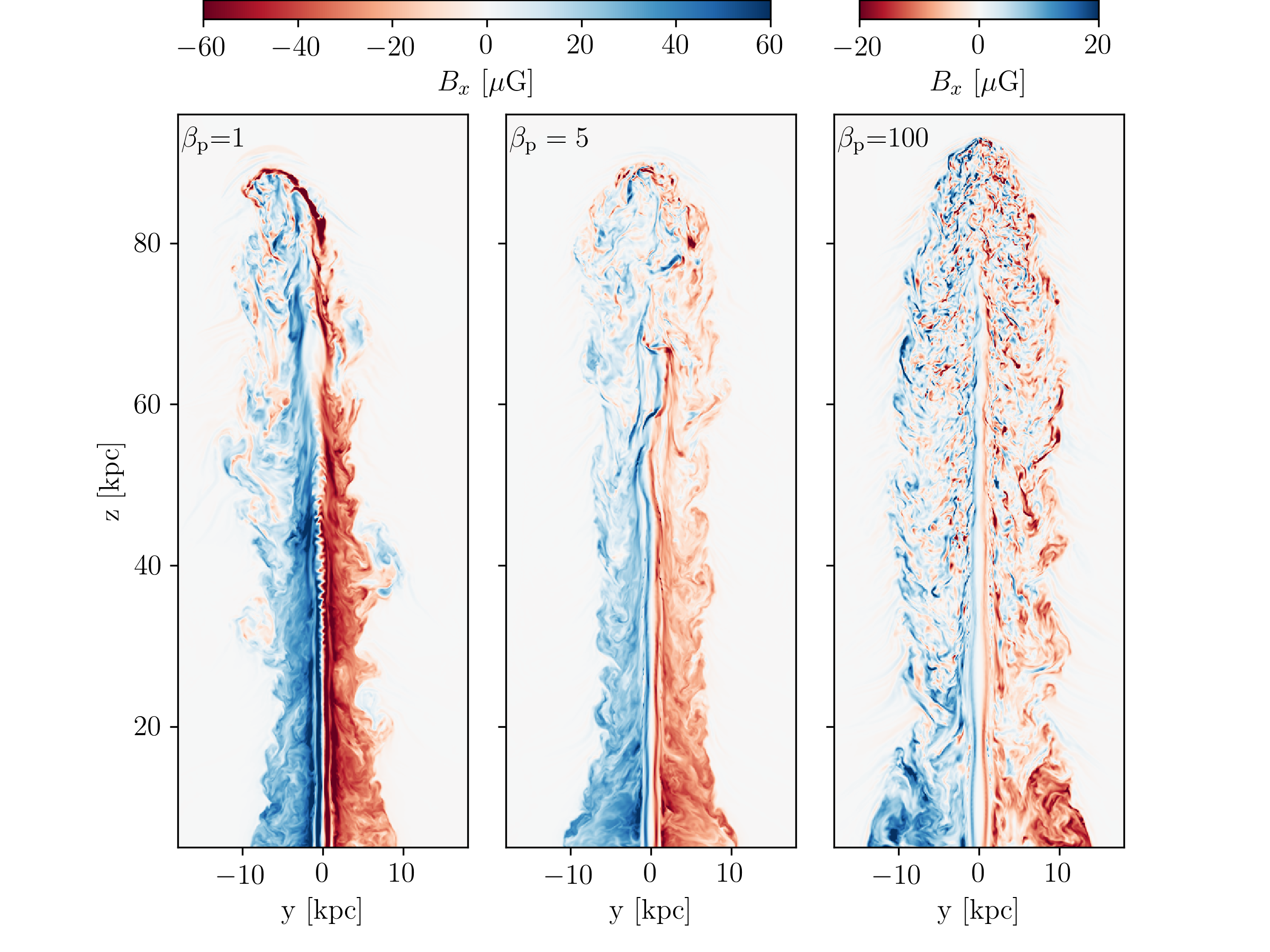}
     \caption{Slices (in the $y-z$ plane) of the x-direction component distribution of the magnetic field for models A, B, and C, respectively.}
     \label{fig5}
   \end{center}
 \end{figure*}

\fullfigref{fig5} shows the slices of the x-direction component of the magnetic field for models A, B, and C, respectively.
While the jet beam is injected with a pure toroidal magnetic field, the inverse field is randomly distributed in the cocoon.
The toroidal magnetic field in the cocoon is weaker, i.e., closer than $z < 40$ kpc.
Further, reversing fields are dissipated in the cocoons of models A and B at $z < 40$ kpc.
We describe the field structures in the beams in more details in \secref{ch5_sec:dissipation}.

\begin{figure*}
  \begin{center}
     \includegraphics[width=1.0\textwidth]{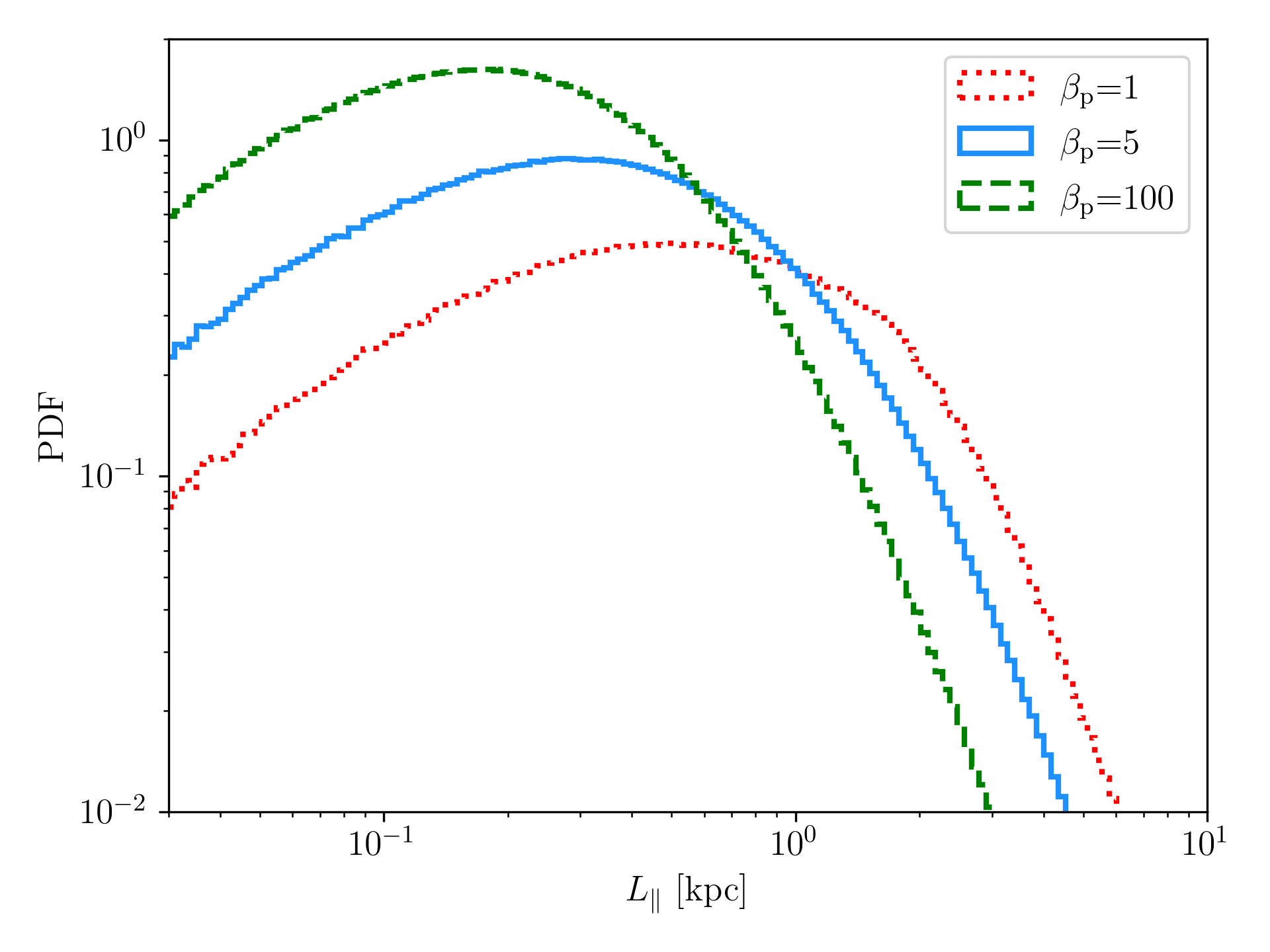}
     \caption{Probability distribution functions of the characteristic length scale parallel to the magnetic field in the cocoon  where $z > 40$ kpc for model A (red dotted), B (blue solid), and C (green dashed) at the end of the simulations. We define the cocoon as grids with the electron temperature higher than $10^8$ K and $z >$ 40 kpc.}
     \label{fig6}
  \end{center}
 \end{figure*}

A magnetic filament develops in the cocoon at $z>40$ kpc in model A, because the strong initial toroidal magnetic field that flows down with the countercurrent creates sufficient magnetic tension to suppress turbulent motion.
The typical length of the filament is several kpc, longer than the filament in model B.
Due to shock compression, filaments are formed around the jet head, and have stronger magnetic fields than the injected ones.
In contrast, for the cocoon of model C, small-scale turbulence is excited.
To discuss the length-scale of the magnetic filament quantitatively, we evaluate the length-scales parallel to the magnetic field as \citep{2004ApJ...612..276S,2011ApJ...739...82B,2020MNRAS.499..681M}
\begin{equation}
    L_{\parallel} = \sqrt{ \frac{|\bm{B}|^4}{|(\bm{B}\cdot \nabla) \bm{B}|^2}    }.
\end{equation}
\fullfigref{fig6} shows the volume-weighted probability distribution function (PDF) of the $L_{\parallel}$ in the jet cocoon where the electron temperature is higher than $10^8$ K and $z > 40$ kpc.
It can be seen from the PDF that the typical value of the length scale increases with stronger magnetization.
The cocoon of model C is filled in smaller vortices, whose typical length scale is about 0.3 kpc.
The volume occupations of the length scales that are longer than 1 kpc are 6.5, 16.5, and 28.0 \% for models A, B, and C, respectively.
These trends are consistent with the results of previous MHD studies \citep{2020MNRAS.499..681M}.


\subsection{Temperature distribution}
\begin{figure*}
  \begin{center}
     \includegraphics[width=1.0\textwidth]{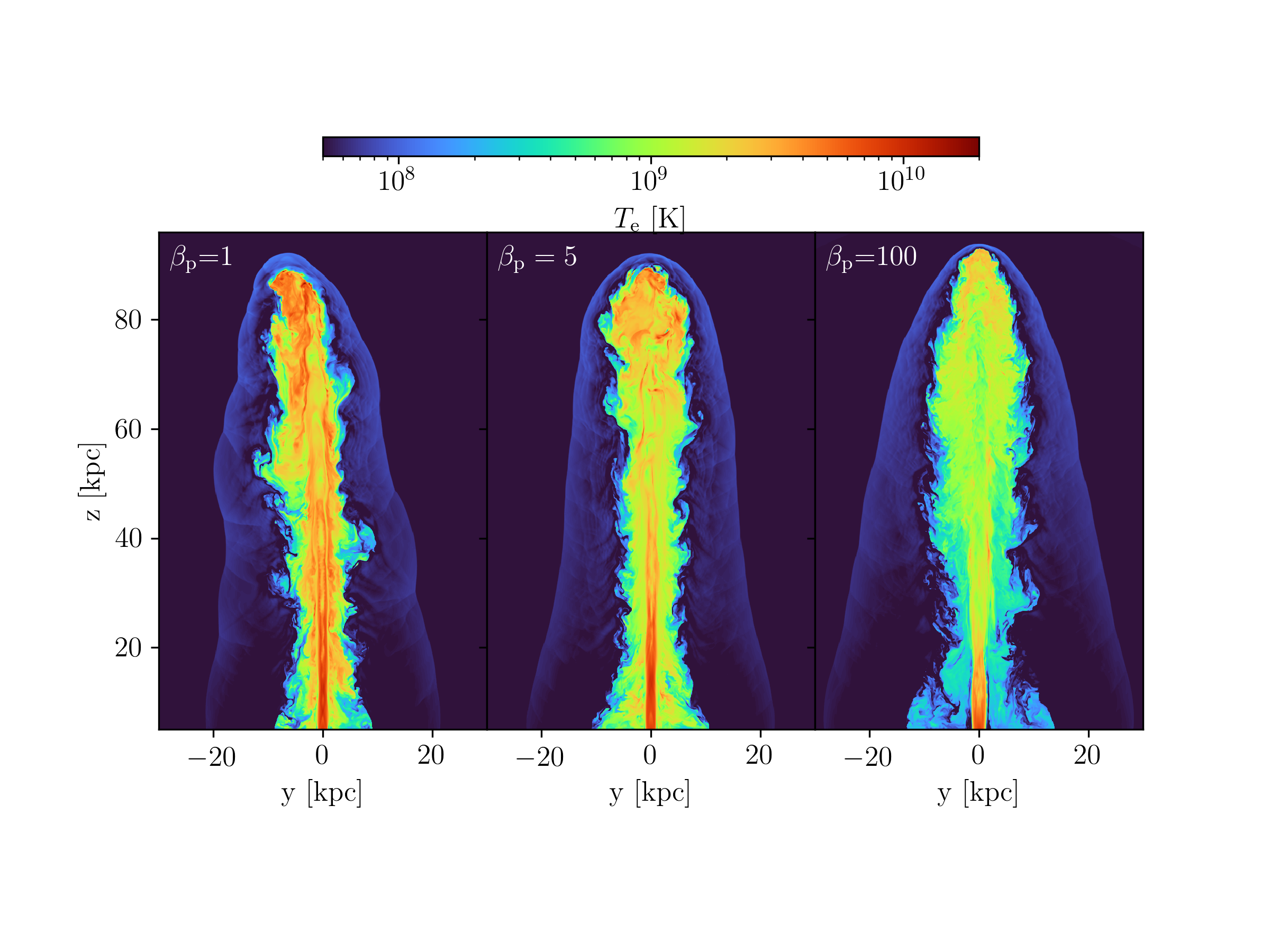}
     \caption{Slices (in the $y-z$ plane) of the electron temperature distribution for models A, B, and C, respectively.}
     \label{fig7}
  \end{center}
 \end{figure*}

\fullfigref{fig7} shows the distribution of the electron temperature at the $yz-$plane for models A, B, and C.
At first glance, the electron temperature in the jet is proportional to the strength of injected magnetic field.
This implies that the sub-grid for turbulence heating plays an important role in the evolution of the electron temperature (see also \secref{ch5_sec:lobe}).
Here, we recall that the electron heating fraction for turbulence is proportional to inverse plasma beta $\beta\rmf{p}^{-1}$.

Thermal electrons propagating through the beam are not subject to the heating energy of the internal shock, but the electrons are heated in the jet termination region.
Subsequently, hot electrons are stored in the cocoon.
Although this physical image  is similar to the result of the two-dimensional case (see Figure 1 in \citet{2020MNRAS.493.5761O}), the difference between them is that the turbulence heating also works in the beam.
Thus, the electrons are heated locally in the beam for models A and B.

\begin{figure*}
   \begin{center}
     \includegraphics[width=1.0\textwidth]{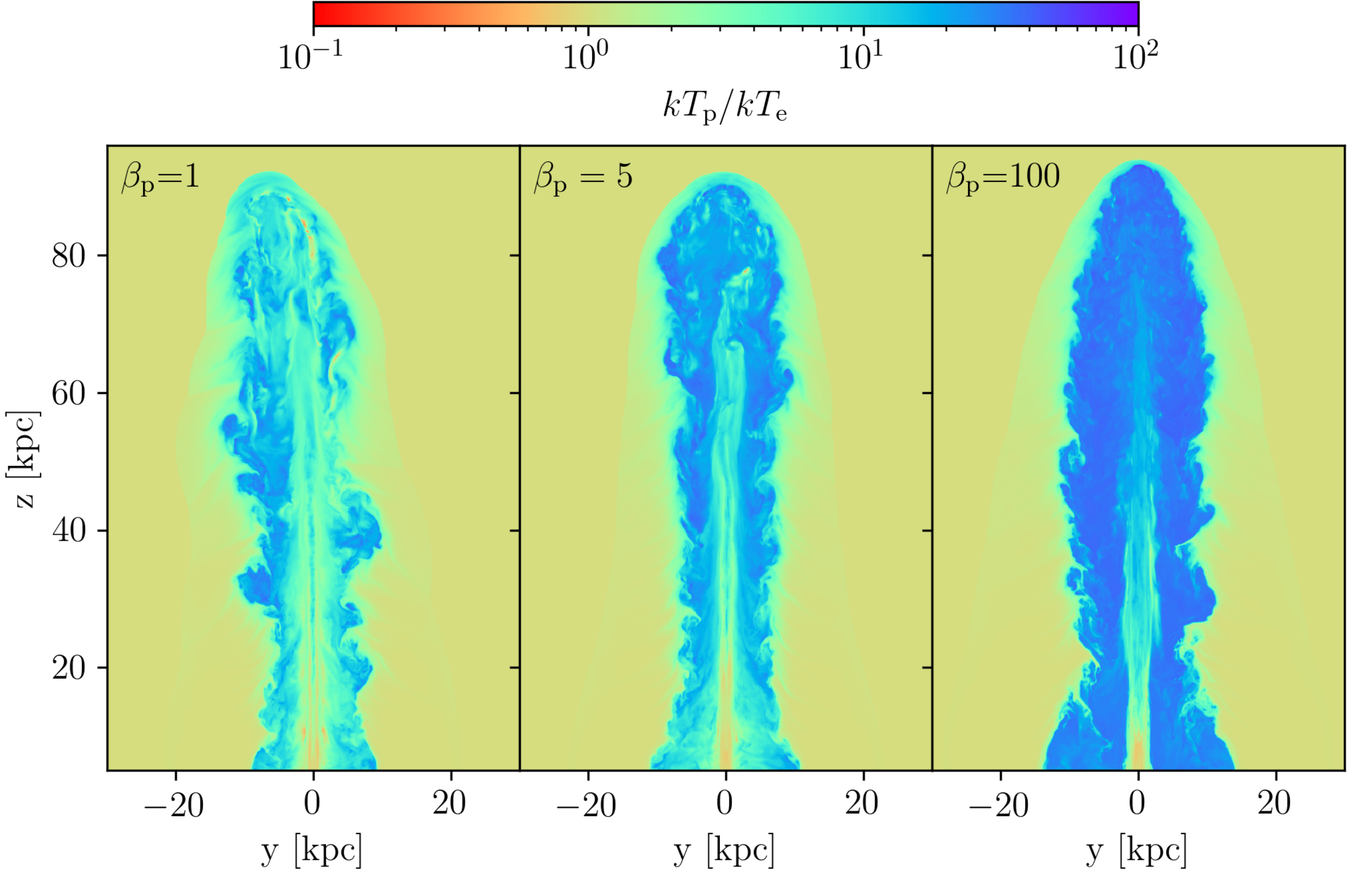}
     \caption{Slices (in the $y-z$ plane) of the ratio of proton to electron temperature distributions for models A, B, and C, respectively.}
     \label{fig8}
   \end{center}
 \end{figure*}

In contrast to electrons, protons receive most of the shock heating, as we set $f\rmf{e} = 0.05$.
Thus, proton temperatures are several ten times higher than electron temperatures in the cocoons for all models (\figref{fig8}).
Electrons are in the relativistic temperature in range of $T\rmf{e} \sim 10^{9} - 10^{10}$ K.
Hotter electrons are located along magnetized filamentary structures formed by shock compression for models A and B.
In particular, electron temperatures are higher than the protons' temperature in some filaments of model A.
Meanwhile, for model C, the distribution of the ratio of the proton to electron temperature is monochromatic, $kT\rmf{p}/kT\rmf{e} \sim 40$, in the cocoon.

\subsection{Lobe energetics} \label{ch5_sec:lobe}
\begin{figure*}
   \begin{center}
     \begin{minipage}{0.49\hsize}
        \includegraphics[width=1\textwidth]{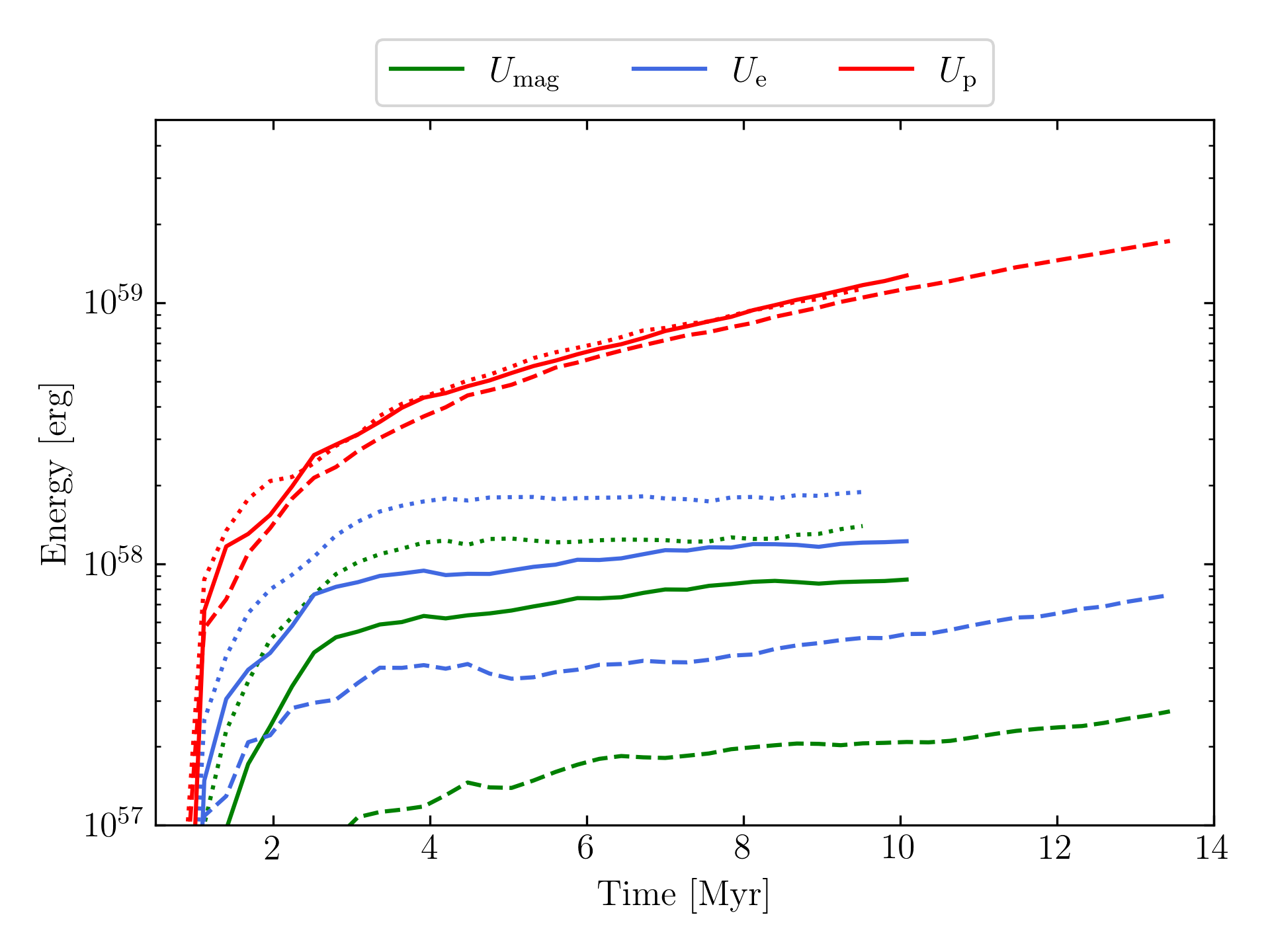}
     \end{minipage}
     \begin{minipage}{0.49\hsize}
        \includegraphics[width=1\textwidth]{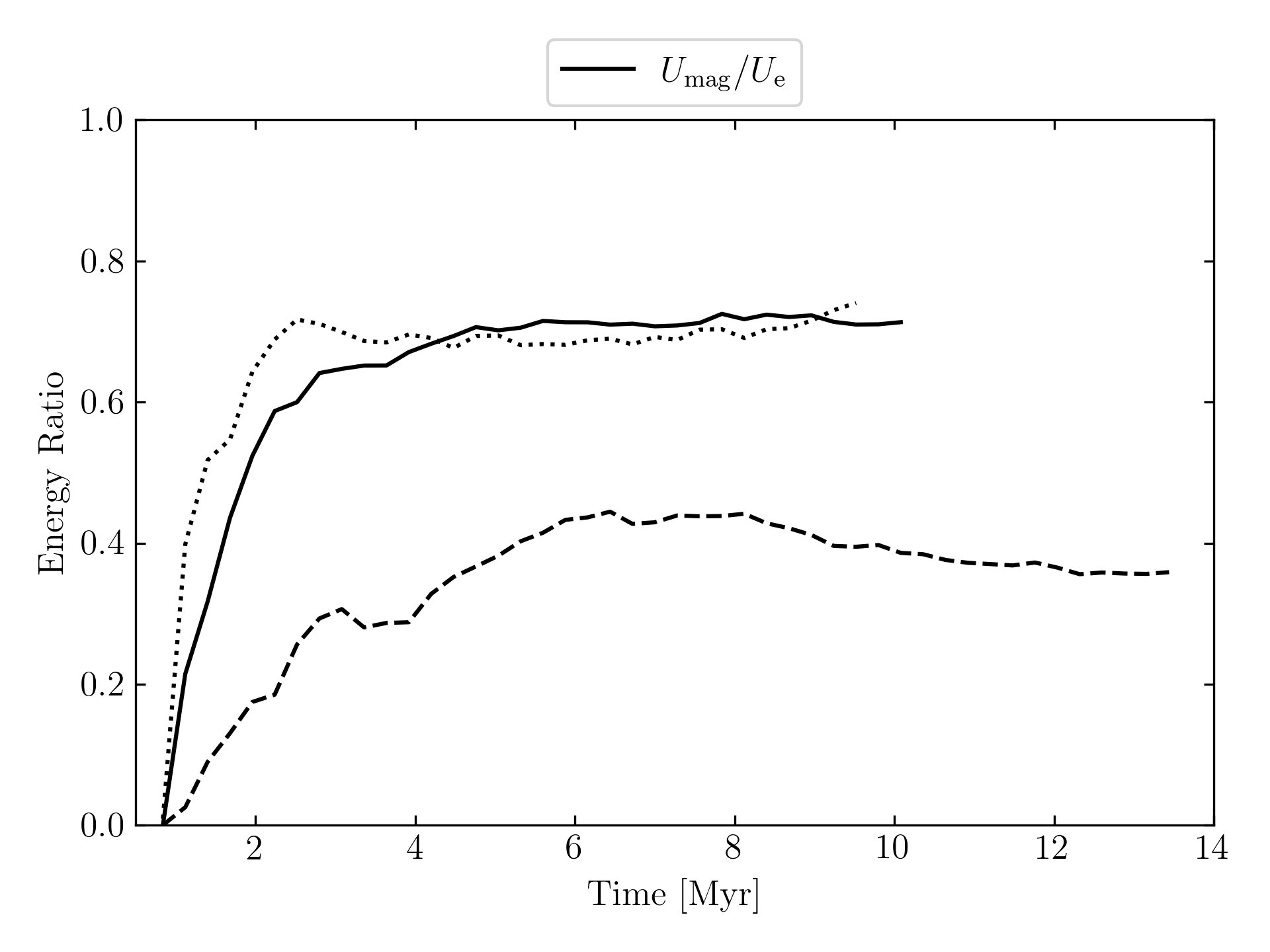}
     \end{minipage}
     \caption{ Energies in the cocoon as a function of time for all models.
     {\bf Left}: Time evolution of different energy components of the cocoon for model A (dotted lines), B (solid lines), and C (dashed lines), respectively. We define the cocoon as grids with an electron temperature higher than $10^8$ K. {\bf Right}: Time evolution of the ratio between the magnetic field and the electron energy for model A (dotted lines), B (solid lines), and C (dashed lines).}
     \label{fig9}
   \end{center}
 \end{figure*}

The left panel of \figref{fig9} displays the time evolution of different energy components of the cocoon for all models.
Proton thermal energy is the dominant energy component of the cocoon ($U\rmf{p} \gg U\rmf{e}$), i.e., cocoons are supported by the proton pressure (see red and blue lines in the left panel of \figref{fig9}).
We confirm that the kinetic energy is comparable to the proton thermal energy, and that 20 percent of total injected energy is converted to the ICM at $t = 10$ Myr.
Because the electron heating fraction of turbulence is an increasing function of $\beta\rmf{p}^{-1}$, there is a positive correlation between the electron thermal energy and field strength (see dashed lines, solid lines, and dotted lines for magnetic energy and electron thermal energy in the left panel of \figref{fig9}).

In the right panel of \figref{fig9}, we plot the time evolution of the ratio between the magnetic and electron thermal energies in the cocoon.
The energy ratio for models A and B saturates at $\sim 0.7$.
The electron pressure is described by $p\rmf{e} = (\gamma\rmf{e}-1) u\rmf{e} = 2u\rmf{e}/3$ for $\gamma\rmf{e} = 5/3$ (from \eqref{ap_eq:eos}), and hence it is in pressure equilibrium with the magnetic field for model A and B.
Meanwhile, in the case of model C, the electron pressure is larger than the magnetic pressure.

\begin{figure*}
   \begin{center}
     \includegraphics[width=1.0\textwidth]{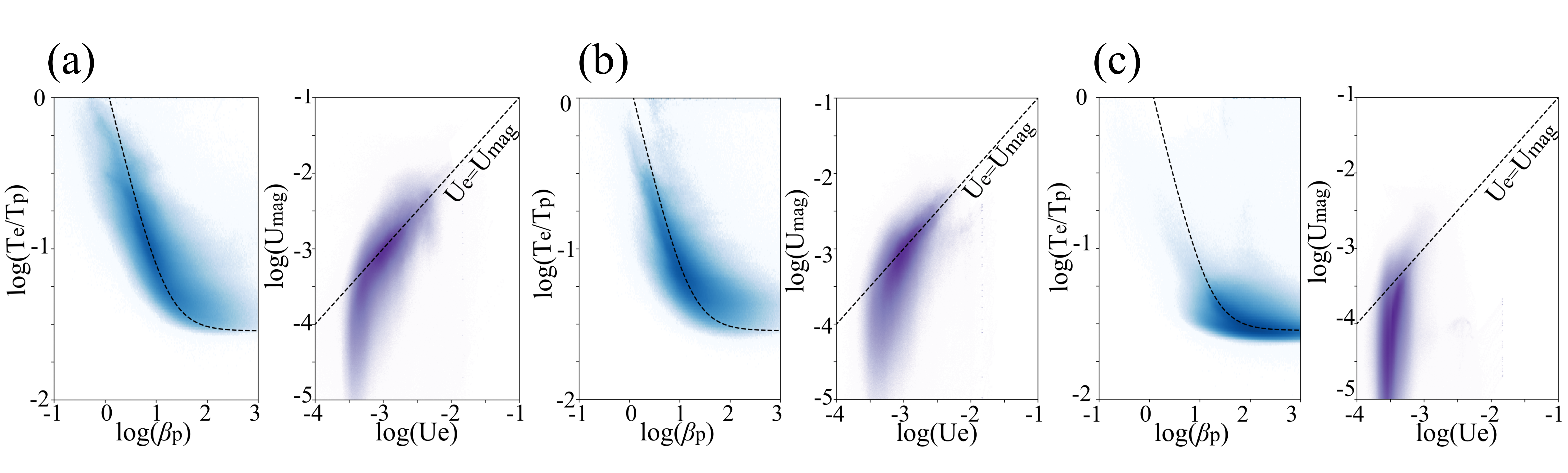}
     \caption{ Plots related to relationship of three energy components -- electron thermal energy, proton thermal energy, and magnetic field energy.
     {\bf Left}: $T\rmf{e}/T\rmf{p} - \beta\rmf{p}$ histogram for regions in the cocoon for model A (a), model B (b), and model (c) at $ t = 9.52~9.94$ and 13.02 Myr, respectively. The dashed line depicts the electron to proton temperature ratio corresponding to the equilibrium state for plasma $\beta\rmf{p}$, as implied by the turbulence heating in equation \ref{ch5_eq:turb_equi}. {\bf Right}: Same as left panel, but displaying $U\rmf{e}-U\rmf{mag}$ histogram. Dashed line plots $U\rmf{e} = U\rmf{mag}$.}
     \label{fig10}
   \end{center}
 \end{figure*}

Thermal electrons evolve while their thermal energy is added by a large amount of dissipated energy due to shocks and turbulence.
If the gas reaches a turbulence equilibrium discussed in \citet{2017MNRAS.466..705S}, the final temperature ratio is determined by the electron heating model, which is given by
\begin{equation}
   \left.\frac{T\rmf{p}}{T\rmf{e}} \right|\rmf{eq} = f\rmf{e,turb}(T\rmf{e}, T\rmf{p}, \beta\rmf{p}),
   \label{ch5_eq:turb_equi}
\end{equation}
where we adopt $\gamma\rmf{p} = \gamma\rmf{e} = 5/3$.
For model Aand B, a quasi-steady state of MHD turbulence is observed in the $T\rmf{e}/T\rmf{p} - \beta\rmf{p}$ histogram (left panel of \figref{fig10}ab).
This histogram plots the gas stored in the cocoons.
We observe that the gas distribution follows along the dashed line, which is plotted as equation \ref{ch5_eq:turb_equi}.
This indicates that energy components ($U\rmf{p}$, $U\rmf{e},$ and $U\rmf{mag}$) evolve following the electron heating model of MHD turbulence.
Therefore, the electron heating model of turbulence plays a significant role in the determination of the gas thermal evolution in our models.
The heating ratio of protons to electrons, $Q\rmf{p}/Q\rmf{e}$ in the turbulence heating model saturates at $\sim 30$ for $\beta\rmf{i} > 10$, and therefore the minimum temperature ratio is located at $(T\rmf{e}/T\rmf{p}) = 1/30$ (see equation \ref{ch5_eq1:kawazura}).
Another view on the turbulence equilibrium is the relationship between $U\rmf{e}$ and $U\rmf{mag}$, shown in the right panel of \figref{fig10}.
The gas distributes along the line, $U\rmf{e} = U\rmf{mag}$, in this histogram.
We confirmed that the volume occupations of the gas where $0.666 <U_{\rm e}/U_{\rm mag} < 3$ are 0.82 and 0.80 for model A and B.
Therefore, the electrons evolve toward energy equipartition (pressure equilibrium) with the magnetic energy when $\beta\rmf{p} < 10$ in the cocoon for models A and B.
Note that The fraction of electron heating become a constant, $f\rmf{e} \sim 1/30$, when the proton plasma-$\beta\rmf{p}$ is higher than 10, i.e., the fraction of electron heating does not depend on the magnetic fields in this regime. 
Thus, a part of gas does not distribute along the line that $U\rmf{e} = U\rmf{mag}$ in a $U\rmf{e}-U\rmf{mag}$ histogram.

For model C, proton plasma-$\beta\rmf{p}$ is higher than 10 across the whole region of the cocoon.
Thus, $U\rmf{e}/U\rmf{p} \sim 1/30$ in the cocoon for model C after 5 Myr, and there is no relation between the electron thermal energy and the magnetic field energy (see \figref{fig10}c).
The volume occupations of the gas where $0.666 <U_{\rm e}/U_{\rm mag} < 3$ for model C are 0.35.
Further, the magnetic energy is subdominant compared with the electron thermal energy.
Notably, $U\rmf{e}/U\rmf{mag}$ saturates at $\sim 0.4$ for model C in the right panel of \figref{fig10}.
However, the saturation value depends on the proton plasma beta when $\beta\rmf{p} \gg 10$ in a cocoon.

\section{Discussion} \label{sec:5}
\subsection{Small-scale dissipation in jet beam} \label{ch5_sec:dissipation}
As we reported in \secref{sec:3}, beams suffer MHD instabilities.
The growth of instabilities leads to the formation of current sheets, where magnetic reconnection takes place.
Magnetic reconnection is a dissipation mechanism that can energize non-thermal particles.
Notably, although we do not explicitly deal with resistivity, magnetic reconnection arises due to numerical dissipation.
The magnetic energy dissipation rate is given by $\eta j^2$, where $\eta$ is the resistivity, and $j$ is the current density.
However, it is difficult to measure the numerical resistivity in ideal MHD simulations.
Thus, following \citet{2017ApJ...835..125Z}, we quantify dissipation to be proportional to the strength of $\bm{j} \cdot \bm{E}$, where $\bm{E} = - \bm{v} \times \bm{B}$ is the electric field.

\fullfigref{append_fig:3} in Appendix \ref{sec:appendD}  displays the volume-renders of a physical quantity $\bm{j} \cdot \bm{E}$ at times that jets reach 60 kpc.
Notably, the color bar scale of the panel (c) is 0.2 times narrower than that of panels (a) and (b).
Peaks of dissipation take place around the jet head for all models due to the shock compression at the termination shocks.
This feature can be regarded as typical for powerful FR-II type jets.
When model A enters the non-linear phase for the kink mode, the beam core is fragmented, which is shown in the beam at $30 < z < 50$ kpc in the left panel of \figref{fig5}.
This fragmented structure identifies as a dissipation region, which shows a small-scale periodic structure at $30 < z < 50$ kpc for model A.
Meanwhile, the value of $\bm{j} \cdot \bm{E}$ is high at the shear layer between the beam and cocoon for model B.
The growth of the Kelvin-Helmholtz mode drive gas mixing between the beam and cocoon in the shear layer \citep{2010MNRAS.402....7M,2020MNRAS.499..681M}.
Therefore, the beam radius of model C in \figref{append_fig:3} looks larger than that of models A and B.

After the jets propagate at 95 kpc, the dissipative structures of the three models are significantly different (\figref{append_fig:3}).
We observe the fragmentation structures at $30 < z < 60$ kpc for model A at this time.
The magnetic energy is dissipated in this region, and hence the flow is laminar downstream of it ($60 < z < 70$ kpc).
Dissipative spots are formed by shock, in particular, and the shock is induced by magnetic pinching at $z = 70$ kpc.
For model B, we observe the abrupt change in the flow direction by the development of a beam kink at $z =$ 60, 70, 80 kpc.
Hence, these localized spots, where the flow is shocked and bent, could be reconnection layers, where efficient particle acceleration would take place.
Such beam disruption can explain the formation mechanism of double hotspots in the western lobe of Cygnus A \citep{1996A&ARv...7....1C} and of multiple knots observed 3C273 \citep{2006ApJ...648..910U}.
Meanwhile, the model C jet does not yield the formation of multiple hotspots.
Although the model C jet has the dissipative spot at the jet head, the dissipation ratio of the magnetic energy is uniformly distributed.
Therefore, the jet at the later phase of model C has a feature of FR-I type jets such as M87, which has an extended diffusive radio lobe without a hotspot \citep{2011MNRAS.417.2789L}.
Jet deceleration by the development of Rayleigh-Taylor instabilities is a possible explanation of FR dichotomy, consistently with previous simulations \citep{2016A&A...596A..12M,2020A&A...642A..69R}.

\subsection{Comparison with observations} \label{sec_ch5:dis_comp_obs}
We discuss that our two-temperature MHD models connect to the observational results.
In particular, we focus on the relationship between the jet mechanical power and radio power.
Observational data sets of X-ray and radio properties are adopted from the PhD thesis by Laura {B{\^\i}rzan} \citep{birzan_phd_thesis} and \citet{2006ApJ...652..216R}.

\subsubsection{Radio power}
The radio power is obtained by the sum of the radio emissivity for the synchrotron emission in the optical thin limit, as follows:
\begin{equation}
    P_{\rm radio} = \int j_{\nu}~dV d\nu,
\end{equation}
where $j_{\nu}$ and $\nu$ are the synchrotron emissivity and the observed frequency, respectively.
We use the formula of the synchrtoron emissitity provided in \cite{1970ranp.book.....P}.

We assume a single power-law distribution for non-thermal electrons, $dN/d\gamma = N_0 \gamma^{-s}$, where $s$, $N_0$ and $\gamma$ are the electron energy index, number of non-thermal electrons, and Lorentz factor of non-thermal electrons.
Meanwhile, the two-temperature MHD simulations account only for the evolution of thermal electrons.
Because previous studies suggest that the non-thermal electron energy is proportional to the thermal gas energy or magnetic energy as a first-order approximation, this approximation is adopted to estimate the radio power from numerical simulations \citep[e.g.,][]{1995ApJ...449L..19G}.
We follow this approximation in the current study.
We adopt the two-type model for approximation of non-thermal electrons as Cases 1T and 2T.
Case 1T is $N_0 = C_0 u\rmf{p}$, and Case 2T is $N_0 = C_0 u\rmf{e}$.
Here, $C_0 = \eta (s-2)(m\rmf{e}c^2)^{-1} \left( \gamma_{\rm min}^{2-s} - \gamma_{\rm max}^{2-s} \right)^{-1}$, and $\eta = 0.2$ is a parameter, which is the ratio of the non-thermal electron energy density to the electron thermal energy density.
Furthermore, we set $\gamma_{\rm min}=100$ and $\gamma_{\rm max} \to \infty$.
Case 1T is corresponds to the single-temperature model, i.e., $T\rmf{e} = T\rmf{p}$.
This model is motivated by the prior works demonstrating the radio emission from AGN jets on assuming thermal equilibrium between protons and electrons \citep[e.g.,][]{1995ApJ...449L..19G}.
We assume that an electron energy index, $s$, is 2.05.
For all calculations, the viewing angle, which is in respect to the $z-$axis, is 80 degrees, and the observed frequency is 144 MHz.

We list the calculated radio powers in \tabref{ch5_tab:sim_radio_mechanical} (see also \figref{fig12}).
The radio power of model A-1T, which is most prominent model, is two orders of magnitude higher than that of model C-2T.
The 1T models have same order of non-thermal electron energy, because the proton temperatures have roughly the same value in lobes (see left panel of \figref{fig9}).
Notably, 1T models assume that non-thermal electron energy is proportional to the proton temperature.
Meanwhile, electron temperatures vary for the three simulation models, they are proportional to the inverse the proton plasma-$\beta$.
Thus, there are a large scatter in the radio powers of 2T models, larger than that of 1T models.
This indicates that the radio power of two-temperature models is sensitive to the magnetic field energy.

\begin{table}
  \begin{center}
  \caption{Radio power and the amount of PdV work for the simulation results}
  \label{ch5_tab:sim_radio_mechanical}
    \begin{tabular}{c c c} \hline
      Model & $P\rmf{radio}$               & $pV$                   \\
            & $[10^{42}~{\rm erg~s^{-1}}]$ & $[10^{58}~{\rm erg}]$  \\ \hline\hline
      A-1T  & 63.2 & 5.95 \\
      A-2T  & 7.23 &  -   \\
      B-1T  & 52.2 & 6.58 \\
      B-2T  & 2.83 &  -   \\
      C-1T  & 24.0 & 9.22 \\
      C-2T  & 0.62 &  -   \\ \hline
    \end{tabular}
\end{center}
\end{table}

\subsubsection{X-ray cavity and jet mechanical power}
Because our simulation assumes a constant energy input corresponding to the active phase of the jet, we can calculate the true mechanical power in units of erg/s.
At the same time, we obtain snapshot quantities and the gas pressure of surrounding ICM through the actual observations.
Therefore, the same observation method is adopted in this work.
Meanwhile, mechanical power is measured using PdV work, as follows in X-ray observation.
\begin{equation}
  P\rmf{cav} = 4p\rmf{gas}V t\rmf{age}^{-1},
\end{equation}
where $t\rmf{age}$ is the outburst age of jets.
Lobes and ICM are approximately in the pressure equilibrium state.
Thus, we use the initial ICM pressure around middle of lobe $p\rmf{gas} \sim ~ 2.0 \times 10^{-10} ~{\rm erg~cm^{-3}}$ to calculate the cavity power (see \figref{fig1}).
Three estimations are commonly used for the outburst age: the buoyancy time $t\rmf{buoy}$, refill time $t\rmf{r}$, and sound crossing time $t\rmf{c}$, generally $t\rmf{c} < t\rmf{buoy} < t\rmf{r}$ \citep{2007ARA&A..45..117M}.
It is appropriate to use the sound crossing time in our model, because jets are in an active phase.
The cavity volume $V$ is calculated by integrating numerical grids with an electron temperature exceeding $10^{8}$ K.
We confirm that the region whose electron temperature exceeds $10^{8}$ K has a lower density than ICM.

\begin{figure}
  \begin{center}
    \includegraphics[width=1\columnwidth]{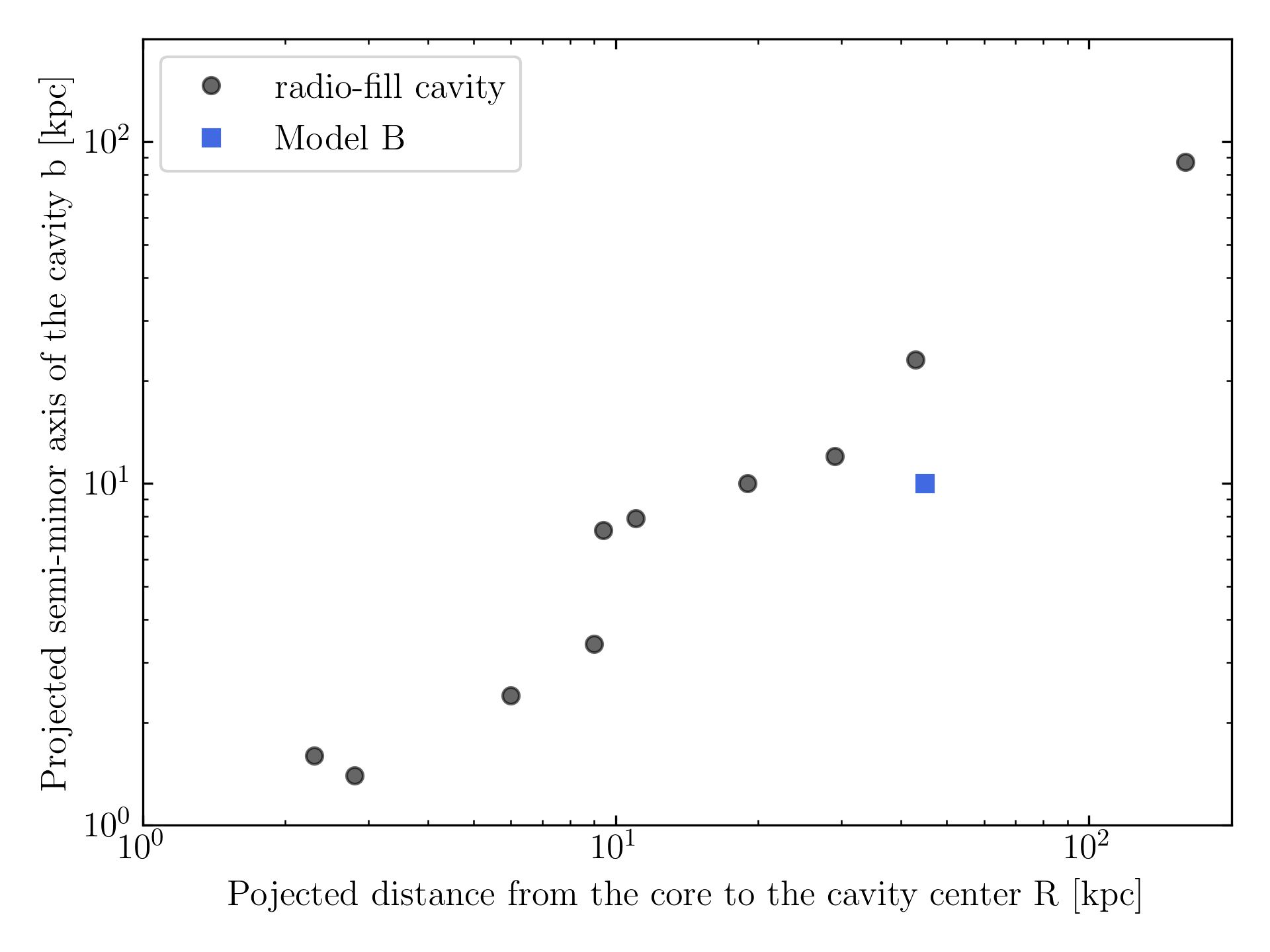}
    \caption{Projected distance from core to cavity center $R$ vs. Projected semi-minor axis of cavity $b$. Black circles show the radio-filled cavities taken from \citet{2006ApJ...652..216R}. Blue square shows our result for model B at $t = 9.94$ Myr. Because $R$ and $b$ for models A and C have similar values as model B, we do not show them. }
    \label{fig11}
  \end{center}
\end{figure}

First, we mention the morphological properties of the X-ray cavity.
In \figref{fig11}, we compare our simulation of model B with observations about the relationship between the projected distance from the core to the cavity center $R$ and the projected semi-minor axis of the cavity $b$.
Because $R$ and $b$ for model A and model C have similar values as for model B, we only show the result of model B in \figref{fig11}.
The axis of the cavity of model B is roughly consistent with the observation, though the cavity is narrow compared with observed ones.
This long and narrow structure is characteristic of a powerful (kinetic-dominated) jet, observed in numerous previous studies \citep[e.g.,][]{2016A&A...596A..12M,2019MNRAS.482.3718P,2020MNRAS.499..681M}.
To create a broad cavity, the jet must propagate slowly to have sufficient time to expand.
Thus, a simple solution to forming a broad cavity is to model a low-density jet \citep{2021ApJ...910..149O} and/or a low-power jet \citep{2020MNRAS.499..681M}.
Otherwise, a long-periodic precession may play an important role in forming observed broadened cavities \citep{2020MNRAS.499.5765H}.
The jet for model B decelerates and has precession due to the development of large-scale kink modes.
However, the speed of the lateral expansion is also decelerated to about the sound speed of ICM.
Therefore, we cannot expect the axis ratio ($R/b$) to decreases at later times \citep{2020MNRAS.499..681M}.

Next, we discuss the energy and age by the observational method.
The cavity energies are calculated as $E\rmf{cav} = 4p_{\rm gas}V = 2.4 \times 10^{59},$ $2.6 \times 10^{59},$ and $3.7 \times 10^{59}$ ${\rm erg}$ for model A, B, and C, respectively.
Further, the sound crossing time $t\rmf{c} = R/c_{\rm s,ICM}$ is 51.6 Myr, where we adopt that $R = 45$ kpc and $c\rmf{s} = 830$ km/s (corresponding to ICM temperature $T = 5$ KeV).
Therefore, we estimate the mechanical power $P\rmf{cav} = 1.5 \times 10^{44},$ $1.6 \times 10^{44},$ and $2.3 \times 10^{44}$ ${\rm erg \ s^{-1}}$ for model A, B, and C, respectively.
Hence, the actual age from our simulations is $\sim 10$ Myr for all models, and the sound crossing time underestimates the mechanical power by a factor $\sim 5$.
Even if we estimate the mechanical power using the simulation time, the injection energy of the jet is $\sim 10$ times higher.
The reason is attributed to the conversion of the jet energy to the thermal energy of ICM through shocks and sound waves.
Furthermore, the cocoon of model B is still over-pressured with respect to the ICM while the mechanical power is measured under the assumption that the radio lobe and ICM have reached in the pressure equilibrium state.

\subsubsection{Relationship between radio power and mechanical power}
\begin{figure}
  \begin{center}
    \includegraphics[width=1\columnwidth]{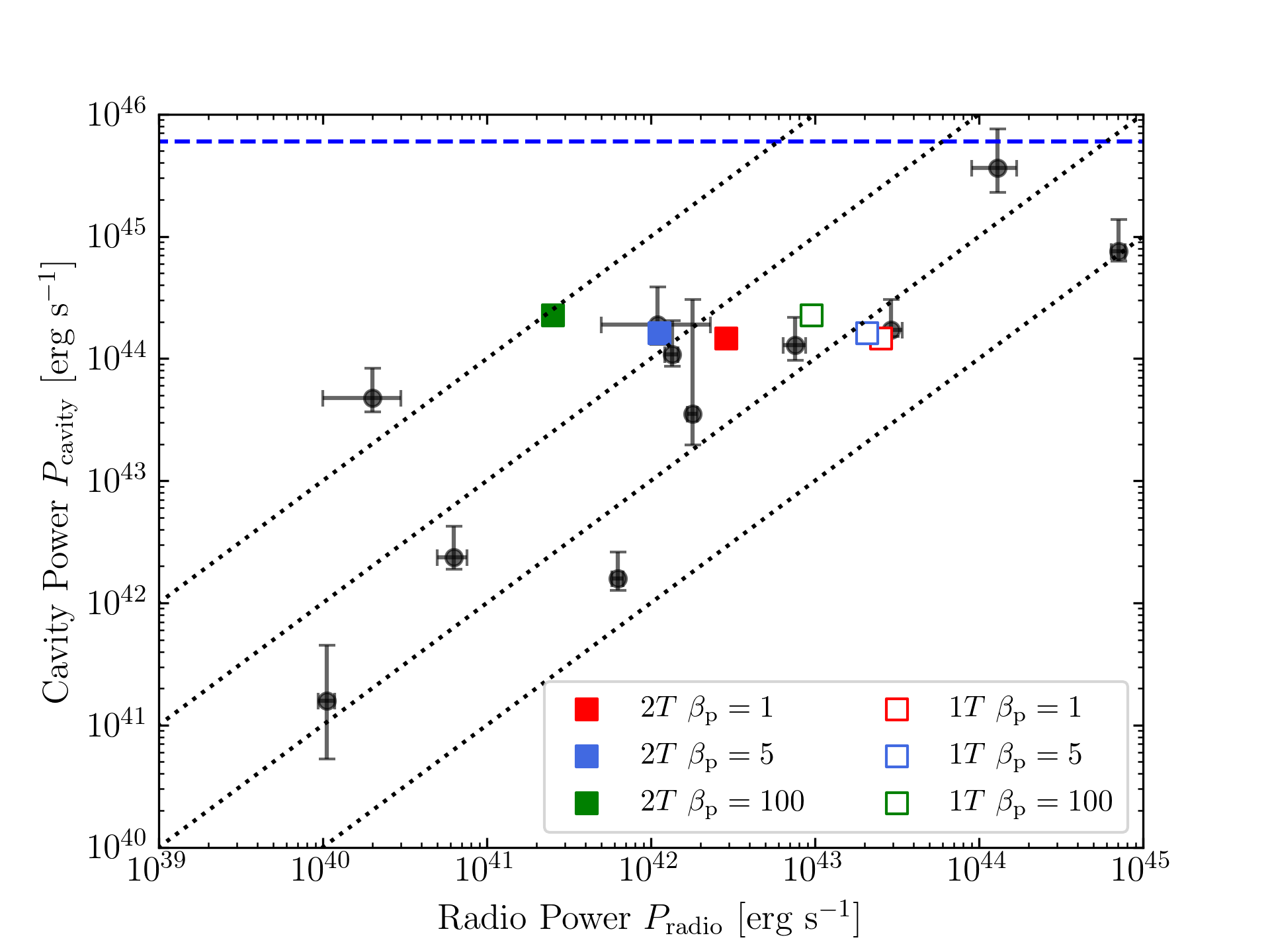}
    \caption{Radio synchrotron power $P\rmf{radio}$ versus jet mechanical power estimated from X-ray cavity system $P\rmf{cav} = 4pV t^{-1}\rmf{c}$ \citep[adopted from Laura {B{\^\i}rzan}'s PhD thesis ][]{birzan_phd_thesis}. Filled black symbols show radio-filled cavities (which include intermediate cases). Symbols and wide error bars denote the values of the mechanical power calculated using the sound speed. Open squares and filled squares show our results of Cases 1T and 2T for models A (red), B (blue), and C (green), respectively.
    The diagonal dotted lines (dashed lines) represent ratios of constant mechanical power to radio luminosity.
    Blue dashed lines depict the total injection energy of our jet model, given by equation \ref{ch5_eq:jet_power}.}
  \label{fig12}
  \end{center}
\end{figure}

The plot of the jet mechanical power $P\rmf{cav}$ versus the synchrotron radio power $P\rmf{radio}$ is shown in \figref{fig12}.
It provides physical insights for jet energetics, including non-radiating proton thermal energy.
Naively, protons can be energetically dominant over radiatively inefficient lobes ($P\rmf{cav} \gg P\rmf{radio}$), as the pressure of non-radio emitting protons supports the expansion of the cocoon.
Our jets are active during simulation time.
Therefore, we compare our results with radio-filled cavities, except for radio-ghost cavities.
Although we plot the radio-filled cavities including the intermediate cases in \figref{fig12}, the samples have a large scatter in the relationship, $P\rmf{cav}/P\rmf{radio} \sim 1 - 1000$.

We find that our two-temperatures model explains radiatively inefficient lobes.
In Case 2T of all models, the lobes tend to be more radiatively inefficient than those of Case 1T.
Because the electrons lack thermal energy compared with Case 1T, the radio powers are weak.
Meanwhile, protons have a large contribution for the cavity power $P\rmf{cav}$.
Thus, radiative efficiencies, $P\rmf{cav}/P\rmf{radio}$, for Case 2T are 10-30 higher than those for the 1T case.
The ratio of the radio power between models A and C for Case 2T is higher than that for Case 1T.
This difference is attributed to electron heating being coupled with the strength of magnetic fields (see \secref{ch5_sec:lobe}).

This result indicates that the pure protons-electrons jet has difficulty to form radiatively efficient lobes, such as Cygnus A, which is located at $P\rmf{cav}/P\rmf{radio}=1$ in \figref{fig12}, without exotic processes.
One of the exotic processes, herein, is the efficient acceleration for electrons.
To create a radiation-efficient lobe, an acceleration mechanism of $\eta \gg 10$ is needed, where the energy of non-thermal electrons is at least an order of magnitude larger than that of the thermal ones.
Alternatively, thermal electrons are dominant over protons, as they are efficiently heated by shocks and turbulence.
Nevertheless, we have not observed this image from several PIC simulations of shock and turbulence
\citep[e.g.,][]{2019MNRAS.485.5105C,2019PhRvL.122e5101Z,2020arXiv200712050Z}.
Another possibility to form radiatively efficient lobes is a strong magnetic field of jets, $\beta_{\rm p} \ll 1$.
However, this possibility is not favored in some observations \citep{2005ApJ...626..733C,2015PASJ...67...77I}.

The existence of a large number of positrons could likewise explain the radiatively efficient lobes.
Our simulations model the pure electron-proton jet.
Thus, to achieve $P\rmf{cav}/P\rmf{radio} = 1$ in our simulations of models A and B (Case 2T), the number density of leptons must be at least a hundred times larger than that of protons, because $P\rmf{radio}$ is proportional to their number density.
Notably, the plasma momentum would be represented by protons under this assumption, because the protons mass is a thousand magnitude greater than the lepton mass.
Thus, it is expected that there is no difference in the electron heating process.
Meanwhile, in the case of model C, significant population of the pair-plasma is needed, because $P\rmf{cav}/P\rmf{rad} \sim 1000$.
In this condition, the electron (and positron) heating process would change, and, furthermore, the MHD approximation would not be guaranteed.
Finally, analytical models of electron-positron-proton mixture jets likewise achieved consistent results with observed FR-II radio lobes in studies by \citet{2008ApJ...685..828I}, \citet{2016MNRAS.457.1124K} and \citet{2012ApJ...751..101K}.
However, these models do not consider the electron heating process at turbulence and shocks.
Therefore, construction of the new model for the mixture jet based on two-temperature simulations is necessary.

We discuss only the radio-fill cavities, as our jet is always active during the simulation time.
However, several radio and X-ray observations imply multi-episodic jet activity \citep[e.g.,][]{2007ApJ...659.1153W,2020A&A...634A...9M}.
These images make it difficult to model the radio lobes, and lead to a large scatters in the $P_{\rm cav}-P_{\rm radio}$ relation.
To understand the physical condition of the radio lobes, it is important that we find radio lobes in active.
To this end, the future radio survey by the Square Kilometer Array is of great value for studying the radio lobes.

\subsection{Observational implications}

We showed that electrons remain in a trans-relativistic temperature, i.e., the electron energy ranges over a few MeV.
Although it is quite difficult to obtain the observational signals from these electrons, there are several possibilities.
One possibility is that the high-quality Sunyaev-Zel'dovich (SZ) radio observations are able to obtain information on the pressure of thermal electrons \citep{2005A&A...430..799P}.
However, our results indicate that the thermal SZ signal from the radio lobe would not be able to detect this, because the electron pressure is lower than the pressures of ICM protons and radio lobe electrons in our model.
Another possibility is the Cosmic Micro Background inverse-Compton (CMB-IC) spectrum.
If the energy of thermal electrons exceeds GeV, the thermal CMB-IC spectrum would be observed in X-ray observations.
However, these components are not yet detected in the contest of radio lobes.
In contrast, the thermal electrons whose energies range with a few MeV scatter photons in the infrared and optical range \citep{2002A&A...383..423E}.

\section{Summary} \label{sec:6}
We carried out two-temperature MHD simulations for the three models whose jets have different magnetic fields.
The jets propagated along 90 kpc in the cluster center, whose environment is roughly consistent with the Cygnus cluster.
We study the dynamics and electron heating in the sub-grid scale of the radio lobes, and thus two sub-grid electron heating mechanisms were considered.
Because the jets have both the turbulence and the strong shock waves, we use the sub-grid models at shock waves and at the turbulence in a hybrid manner.

The main findings achieved in this study are listed as follows:
\begin{enumerate}[(i)]
  \item Strongly magnetized jets suffer from the development of a non-axisymmetric, current-driven kink mode.
  Meanwhile, weakly magnetized jets are decelerated by the high-mixing ratio between the jet beam and cocoon gas, which were induced by Rayleigh-Taylor and Kelvin-Helmholtz instability.
  We show that the Alfv\'{e}n crossing time, $\tau_{\rm kink}$, is a good indicator for the time scale of the development of the kink mode, even if the jets have purely toroidal fields at the injected region. 
  \item Electrons heat up at the jet termination region, and hot electrons are stored in the cocoon.
  The electron heating fraction for turbulence is proportional to the inverse plasma beta $\beta\rmf{p}^{-1}$.
  Therefore, a jet with a strong magnetic field has a higher electron temperature in the cocoon.
  \item Small-scale turbulence develops in the weakly magnetized jets. In contrast, the strongly magnetized jets have magnetized filaments, as the magnetic tension suppresses the turbulence motion.
  \item The protons are energetically dominant over the electrons in the cocoon. First, most of the bulk kinetic energy of the jet is converted into thermal energy of protons through shocks. Second, while magnetic fields are relatively strong, shocked-electrons stored in the cocoon evolve toward energy equipartition with magnetic energy through turbulent dissipation.
  \item The strong current is induced by the kink instability. Therefore, high-temperature and high-magnetization multiple hotspots are formed in the beam.
  \item The low-density cavity of our model is narrow compared with observed radio-fill cavities. This suggests that the density of jets is slightly lower than that determined by our model. The propagation velocity is faster than the sound speeds of ICM, and some fractions of the jet energy are converted into the thermal energy ICM.
  Thus, the jet mechanical energy estimated by X-ray observation is 10 times lower than the injected kinetic energy in our simulations.
  \item Radio powers estimated from the electron thermal energy are ten times lower than those estimated from the proton thermal energy, which correspond to the one-temperature approximation. Two-temperature models quantitatively explain the radiatively inefficient lobes ($P_{\rm cav}/P_{\rm radio} \sim 100$). Further, our results indicate that it is difficult to explain radiatively efficient lobes, such as Cygnus A, unless non-thermal energy is more than one order of magnitude larger than the thermal electron energy.
\end{enumerate}

In the current study, we only focused on the results of the property of the jetted plasma.
However, the weak shocks around radio lobes are frequently observed in X-ray observations \citep[e.g.,][]{2018ApJ...855...71S}.
We confirm that shocked-ICM plasma occurs in two-temperature states (see \figref{fig8}), and hence we aim to report thermodynamics and X-ray properties of shocked-ICM in paper II.

%
\begin{acknowledgements}
We thank the anonymous referee for the useful comments that greatly improved the presentation of the paper.
    This work was supported by JSPS KAKENHI Grant Numbers JP22K14032 (T.O.) and 19K03916 (M.M.). Our numerical computations were carried out on the Cray XC50 at the Center
for Computational Astrophysics of the National Astronomical Observatory of Japan. The computation was carried out using the computer resource by Research Institute for Information Technology, Kyushu University.
This work was also supported in part by MEXT as a priority issue (Elucidation of the fundamental laws and evolution of the universe) to be tackled by using post-K Computer and JICFuS and by MEXT as “Program for
Promoting Researches on the Supercomputer Fugaku” (Toward a unified view of the universe: from large scale structures to planets).
\end{acknowledgements}

\bibliographystyle{aa} 
\bibliography{ref} 

\begin{appendix}

\section{Numerical integration} \label{app:1}
In this simulation, we numerically solve equation \ref{eq:conserved}. 
The steps of numerical interaction are as follows:
\begin{enumerate}
  \item Calculate conservation variables from principle variables at the end of the previous time step, and calculate the effective adiabatic index $\gamma\rmf{gas}$ by equation \ref{eq:gamma_gas}.
  \item Adopt an operator split method ($\bm{S} = 0$ in equation \ref{eq:conserved}), and solve the conservative equation (see \citet{2019PASJ...71...83M} for details).
  \item Recalculate conservation variables, except for the gas thermal energy $u\rmf{gas}$, from principle variables. Note that updated electron and proton temperatures are unknown, such that we cannot calculate $u\rmf{gas}$ and $\gamma\rmf{gas}$ at this step.
  \item Solve the entropy equations for electrons and protons (equation \ref{eq:entropy1} and \ref{eq:entropy2}). The detailed procedure is described  in the following section. Then, calculate $u\rmf{gas}$ and $\gamma\rmf{gas}$ using updated electron and proton temperatures.
\end{enumerate}

\subsection{Solve entropy equations} \label{sec:entropy}
We describe the procedure of numerical integration for the entropy equations of electrons and protons.
For simplicity, we describe the method for the one-dimensional coordinate $x$.
First, we need to estimate the energy dissipation at every grid for each time step. 
The dissipation energy is the difference between the thermal energy of the total gas $u_{\rm gas}$, which as obtained from the energy equation, and the sum of the thermal energies that evolved by purely adiabatic evolution for electrons and protons ($u_{\rm e,ad}$ and $u_{\rm p,ad}$).
To compute adiabatic evolution, the right-hand sides of equation \ref{eq:entropy1} and \ref{eq:entropy2} set to zero.
We then solve these equations by the finite-difference method.
Let $x_i$ be the cell center of a uniform grid, $\Delta x$ is the cell width.
Time describes $t^n = n \Delta t$, where $\Delta t$ is the time step.
We adopt third order TVD-Runge--kutta schemes, such that finite-difference equations are given as follows:
\footnotesize
\begin{eqnarray}
  \label{ap_eq:fd_eqs}
  (\rho s)^{(1)}_{i} = (\rho s)^{n}_{i} - \frac{\Delta t}{\Delta x}  [(\rho sv_x)^{n}_{i+1/2}-(\rho sv_x)^{n}_{i-1/2} ], \\
  (\rho s)^{(2)}_{i} = \frac{3}{4}(\rho s)^{n}_{i} + \frac{1}{4} \left[ (\rho s)^{(1)}_{i} - \frac{\Delta t}{\Delta x}  [(\rho sv_x)^{(1)}_{i+1/2}-(\rho sv_x)^{(1)}_{i-1/2} ]\right],\\
  (\rho s)^{n+1}_{i} = \frac{1}{3}(\rho s)^{n}_{i} + \frac{2}{3} \left[ (\rho s)^{(2)}_{i} - \frac{\Delta t}{\Delta x} [(\rho sv_x)^{(2)}_{i+1/2}-(\rho sv_x)^{(2)}_{i-1/2} ]\right],
  \label{ap_eq:fd_eqe}
\end{eqnarray}
\normalsize
where $(1)$ and $(2)$ denote each number of sub-time steps of the TVD-Runge--kutta scheme.
Note that the entropy formulas for electrons and protons are of the same form, and we do not distinguish between them.

According ot \citet{2017MNRAS.466..705S}, we arrange equations \ref{ap_eq:fd_eqs} - \ref{ap_eq:fd_eqe} as follows:
\tiny
\begin{eqnarray}
  \label{ap_eq:s1}
  s^{(1)}_{i} &=& \frac{\rho^{n}_{i}}{\rho^{(1)}_{i}}s^{n}_i - \frac{\frac{\Delta t}{\Delta x} (\rho v_x)^{n}_{i+1/2} }{\rho^{(1)}_{i}} s^{n}_{i+1/2} + \frac{\frac{\Delta t}{\Delta x} (\rho v_x)^{n}_{i-1/2} }{\rho^{(1)}_{i}} s^{n}_{i-1/2} \nonumber \\
  &=& f^{(1)}_i s^{n}_i + f^{(1)}_{i+1/2} s^{n}_{i+1/2} + f^{(1)}_{i-1/2} s^{n}_{i-1/2}.
\end{eqnarray}
\begin{eqnarray}
  s^{(2)}_{i} &=& \frac{3}{4}\frac{\rho^{n}_{i}}{\rho^{(2)}_{i}}s^{n}_i + \frac{1}{4} \frac{\rho^{(1)}_{i}}{\rho^{(2)}_{i}}s^{(1)}_i - \frac{\frac{1}{4}\frac{\Delta t}{\Delta x} (\rho v_x)^{(1)}_{i+1/2} }{\rho^{(2)}_{i}} s^{(1)}_{i+1/2} + \frac{\frac{1}{4}\frac{\Delta t}{\Delta x} (\rho v_x)^{(1)}_{i-1/2} }{\rho^{(2)}_{i}} s^{(1)}_{i-1/2}
   \nonumber \\
  &=& f'^{(2)}_i s^{n}_i + f^{(2)}_i s^{(1)}_i + f^{(2)}_{i+1/2} s^{(1)}_{i+1/2} + f^{(2)}_{i-1/2} s^{(1)}_{i-1/2}.
\end{eqnarray}
\begin{eqnarray}
  \label{ap_eq:s2}
  s^{n+1}_{i} &=& \frac{1}{3}\frac{\rho^{n}_{i}}{\rho^{n+1}_{i}}s^{n}_i + \frac{2}{3} \frac{\rho^{(2)}_{i}}{\rho^{n+1}_{i}}s^{(2)}_i - \frac{\frac{2}{3}\frac{\Delta t}{\Delta x} (\rho v_x)^{(2)}_{i+1/2} }{\rho^{n+1}_{i}} s^{(2)}_{i+1/2} + \frac{\frac{2}{3}\frac{\Delta t}{\Delta x} (\rho v_x)^{(2)}_{i-1/2} }{\rho^{n+1}_{i}} s^{(2)}_{i-1/2}
   \nonumber \\
  &=& f'^{n+1}_i s^{n}_i + f^{n+1}_i s^{(2)}_i + f^{n+1}_{i+1/2} s^{(2)}_{i+1/2} + f^{n+1}_{i-1/2} s^{(2)}_{i-1/2},
\end{eqnarray}
\normalsize
where $f$ and $f'$ are the fractions of the final state represented by the three contributing grids of gas.
When two individual gases are mixed in a constant volume, the total energy is the sum of the initial energies of the gases.
In contrast, the total entropy is not to be the sum of the initial entropy of the gases.
Hence, the finite-volume methods in equations \ref{ap_eq:s1} - \ref{ap_eq:s2} are incorrect.
However, to overcome this problem, we must treat the dynamics of each gas individually.
In this study, according to \citet{2017MNRAS.466..705S}, we solve equations \ref{ap_eq:s1} - \ref{ap_eq:s2} by replacing the entropy with the thermal energy.
For electrons, the relationship between the entropy and the dimensionless temperature is known in equation \ref{eq:ent_temp_invert}.
Then, given the dimensionless temperature, we can calculate the adiabatic index by equation \ref{ap_eq:eos}.
Therefore, the electron thermal energy, $u_{\rm e} = n_{\rm e}kT_{\rm e}/(\gamma_{\rm e}(\theta_{\rm e})-1)$, is a function of the density and entropy:
\footnotesize
\begin{equation}
  u^{n}_{i,i\pm1/2} = u(s^{n}_{i,i\pm1/2}, \rho^{(1)}_{i}),~~ u^{(1)}_{i,i\pm1/2} = u(s^{(1)}_{i,i\pm1/2},~~\rho^{(2)}_{i}),~~u^{(2)}_{i,i\pm1/2} = u(s^{(2)}_{i,i\pm1/2}, \rho^{n+1}_{i}).
\end{equation}
\normalsize
This process corresponds to assuming gas mixing at a constant gas density.
To be safe, we use upwind values of the entropy, $s^{n}_{i+1/2} = {\rm Upwindow} \left( s^{n}\rmf{i}, s^n_{i+1} \right)$ and $s^{n}_{i-1/2} = {\rm Upwindow} \left( s^{n}\rmf{i-1}, s^n_{i} \right)$.
We also calculate the proton thermal energy following the same procedure described above, but protons are non-relativistic in our simulation (see equation \ref{eq:ent_temp_ion}). 
From the above calculation, we can obtain the thermal energy of electron and proton at each Runge-Kutta substeps as
\begin{equation}
u^{(1)}_{i} = f^{(1)}_{i} u^{n}_{i} + f^{(1)}_{i+1/2} u^{n}_{i+1/2} + f^{(1)}_{i-1/2} u^{n}_{i-1/2},
\end{equation}
\begin{equation}
u^{(2)}_{i} = f'^{(2)}_{i} u^{n}_{i} + f^{(2)}_{i} u^{(1)}_{i} + f^{(2)}_{i+1/2} u^{(1)}_{i+1/2} + f^{(2)}_{i-1/2} u^{(1)}_{i-1/2}, 
\end{equation}
\begin{equation}
u^{n+1}_{i} = f'^{n+1}_{i} u^{n}_{i} + f^{n+1}_{i} u^{(2)}_{i} + f^{n+1}_{i+1/2} u^{(2)}_{i+1/2} + f^{n+1}_{i-1/2} u^{(2)}_{i-1/2}.
\end{equation}
Recall that, through above process, the effect of non-adiabatic process is ignored.
Therefore, this is the internal energy that evolved by purely adiabatic evolution, hereafter denoted as $u_{\rm ad}$.

Then, we calculate the dissipation heating at each grid as:
\begin{equation}
  Q^{(1),(2),n+1}_{\rm heat} = u^{(1),(2),n+1}_{\rm gas} - (u^{(1),(2),n+1}_{\rm e,ad}+u^{(1),(2),n+1}_{\rm p,ad}),
\end{equation}
where $dQ_{\rm heat}/dt =  q_{\rm heat}$.
Further, if necessary, the fraction of the electron heating, $f\rmf{e}$, is calculated using MHD quantities.
Dividing the dissipation heat into electrons and protons,
their thermal energies are updated as follows:
\begin{eqnarray}
  u^{(1),(2),n+1}_{\rm e} = u^{(1),(2),n+1}_{\rm e,ad} + f^{(1),(2),n+1}_{\rm e} Q^{(1),(2),n+1}_{\rm heat}, \\
  u^{(1),(2),n+1}_{\rm p} = u^{(1),(2),n+1}_{\rm p,ad} + (1 - f_{\rm e}^{(1),(2),n+1}) Q^{(1),(2),n+1}_{\rm heat}.
\end{eqnarray}
The source term, namely energy transfer via Coulomb and radiative cooling, is updated implicitly by adopting the Newton--Raphson iteration at last sub-step of the TVD-Runge--Kutta scheme.
Finally, we can easily recover the entropy $s^{(1),(2),n+1}$ ( and the temperature $T^{(1),(2),n+1}$ ) for each species using $\rho^{(1),(2), n+1}$ and $u^{(1),(2),n+1}$.

\subsection{Bremsstrahlung radiation cooling}
The bremsstrahlung cooling rate per unit volume for relativistic plasma is given by\citep{1982ApJ...258..335S}
\begin{equation}
  q_{\rm brems} (\theta\rmf{e}, n)  = n^2 \sigma\rmf{T}c\alpha\rmf{f}m\rmf{e}c^2 [F\rmf{ei}(\theta\rmf{e})+F\rmf{ee}(\theta\rmf{e})],
\end{equation}
where $\alpha\rmf{f}$ and $\sigma\rmf{T}$ denote the fine-structure constant and Thomson cross-section, respectively.
In the above equation, $F\rmf{ei}$, and $F\rmf{ee}$ are the dimensionless radiation rates due to proton-electron and electron-electron collisions, respectively.
The approximation formulas of $F\rmf{ei}$, and $F\rmf{ee}$ are respectively
\footnotesize
\begin{equation}
  F\rmf{ei}(\theta\rmf e) =
    \begin{cases}
    4 \left( \frac{2}{\pi^{3}} \right)^{1/2} \theta\rmf{e}^{1/2} (1+1.78 \theta\rmf{e}^{1.34}) & {\rm for}~(\theta\rmf{e} < 1)\\
    \frac{9 \theta\rmf{e}}{2\pi} \left[ \ln{(2\eta\rmf{E}\theta\rmf{e}+0.42)}+\frac{3}{2} \right] & {\rm for}~(\theta\rmf{e} > 1)
    \end{cases},
\end{equation}
\begin{equation}
  F\rmf{ee}(\theta\rmf e) =
    \begin{cases}
      \frac{20}{9\pi^{1/2}} (44-3\pi^2)\theta\rmf{e}^2 \left( 1+1.1\theta\rmf{e}+\theta\rmf{e}^2-1.25\theta\rmf{e}^{2.5} \right) & {\rm for}~(\theta\rmf{e} < 1)\\
      24 \theta\rmf{e} \left[ \ln{2\eta\rmf{E}\theta\rmf{e}}+\frac{5}{4} \right] & {\rm for}~(\theta\rmf{e} > 1)
    \end{cases}.
\end{equation}
\normalsize
Here, $\eta\rmf{E} = \exp{ (-\gamma\rmf{E})}$ and $\gamma\rmf{E} \approx 0.5772$ is Euler's number.

\subsection{Coulomb coupling}
The rate of energy transfer, $q^{{\rm ie}}$, from ions to electrons per unit volume through Coulomb collisions is determined as follows \citep{1983MNRAS.204.1269S,1991ApJ...369..410D};
\footnotesize
\begin{equation}
 \label{eq:qie}
 q^{{\rm ie}} = \begin{cases}
 \frac{3}{2} \frac{m_{\rm e}}{m_{\rm p}} n^2 \sigma_{\rm T} c
 \frac{\ln{\Lambda} (k T_{\rm p} - k T_{\rm e})}{K_2\left(1/\theta_{\rm e} \right)K_2\left( 1/\theta_{\rm p} \right)  } \left[
  \frac{2(\theta_{\rm e} + \theta_{\rm p})^2 +1}{\theta_{\rm p} + \theta_{\rm e}}K_1\left(\frac{1}{\theta_{\rm m}}\right)
  + 2K_0\left(\frac{1}{\theta_{\rm m}} \right)
  \right] & (\theta_{\rm p} > 0.2) \\
 \frac{3}{2} \frac{m_{\rm e}}{m_{\rm p}} n^2 \sigma_{\rm T} c \ln{\Lambda} (k T_{\rm p} - k T_{\rm e}) \frac{\sqrt{\frac{2}{\pi}}+\sqrt{\theta_{\rm p}+\theta_{\rm e}}}{(\theta_{\rm p}+\theta_{\rm e})^{3/2}}  & (\theta_{\rm p} < 0.2),
 \end{cases}
\end{equation}
\normalsize
where $\theta_{\rm m}=\theta_{\rm p} \theta_{\rm e}/(\theta_{\rm p}+\theta_{\rm e})$ and $\theta_{\rm p} \equiv kT_{\rm p}/m_{\rm p}c^2$.
The parameters $\sigma_{\rm T}$, and $c$ denote the Thomson scattering cross-section and the speed of light, respectively.
$\ln{\Lambda}$ is the Coulomb logarithm and approximated to be
\begin{equation}
    \ln{\Lambda} \approx 37.8 + \ln{\left( \frac{T_{\rm gas}}{10^8~{\rm K} } \right)} - 0.5 \ln{\left( \frac{n}{10^{-3}~{\rm cm^{-3}}}\right)}
\end{equation}
for $T > 4 \times 10^{5}~\rm{K}$ \citep{1962pfig.book.....S}.
Functions $K_{0}, K_{1}$, and $K_{2}$ are respectively modified Bessel functions of the second kind of orders 0, 1, and 2.

\section{Shock-finding algorithm} \label{ap_sec:shock_find}
To identify whether the MHD grid is inside the shock zone, we implement shock-finder in MHD code CANS+, and adopt an approach similar to that followed by \citet{2003ApJ...593..599R} and \citet{2015MNRAS.446.3992S}.
Although our method is based on the theory for hydrodynamic shock, the influence of omitting the magnetic field is insignificant for shock-finding.
The inclusion of magnetic fields complicate the system to add two types of compressible shocks, and the measurement of the Mach number for a MHD shock is difficult task.

Here, we use the Cartesian coordinate $(x,y,z)$, and subscript $i \in (x,y,z)$ to denote the direction of each coordinates.
We divide each grid in or out of the shock zone to employ the following criteria
\begin{eqnarray}
  \nabla \cdot \bm{v} < 0,\\
  \nabla T_{\rm gas} \cdot \nabla \rho > 0, \\
  {\mathcal M} > {\mathcal M\rmf{min}},
\end{eqnarray}
where ${\mathcal M} \equiv \sqrt{\mathcal{M}_x^2+\mathcal{M}_y^2+\mathcal{M}_z^2}$ and ${\mathcal M\rmf{min}}$ are the estimated Mach number and a minimum Mach number.
In the simulations, the divergence operator replaces the central differences, and we adopt ${\mathcal M\rmf{min}} = 1.3$.
The Mach number of each grid is estimated from the Rankine-Hugoniot condition across shocks, which is given by \citep{2017MNRAS.465.4500P}
\begin{equation}
  {\mathcal M\rmf{i}}^2 = \frac{1}{\gamma\rmf{gas,1}} \frac{(y-1){\mathcal C}}{{\mathcal C} - [(\gamma\rmf{gas,1}+1) + (\gamma\rmf{gas,1}-1)y ](\gamma\rmf{gas,2}-1) },
\end{equation}
where $y\equiv p\rmf{gas,2}/p\rmf{gas,1}$ and ${\mathcal C}=[(\gamma\rmf{gas,2}+1)y+\gamma\rmf{gas,2}-1]$.
Up and downstream quantities are denoted by subscripts 1 and 2, respectively.
We determine the direction of shock propagation, $d\rmf{s}$, in each grid using the temperature gradient:
\begin{equation}
  d\rmf{s} = - \frac{\nabla T\rmf{gas}}{|\nabla T\rmf{gas}|}.
\end{equation}

\section{Comparison of the electron heating models}
\label{sec:appendC}
We performed the axisymmetric MHD simulations to compare the results by using different electron heating models for MHD turbulence \citep[H10 and K19,][]{2010MNRAS.409L.104H,2019PNAS..116..771K}.
The electron-to-proton heating ratio of H10 is written by
\begin{equation}
  \label{ch5_eq1:howes}
    \frac{Q\rmf{p}}{Q\rmf{e}} = c_1 \frac{c^2_2 + \beta_{\rm p}^{2-0.2 \log_{10}(T_{\rm p}/T_{\rm e}) }}{c^2_3 + \beta_{\rm p}^{2-0.2 \log_{10}(T_{\rm p}/T_{\rm e}) }} \sqrt{\frac{m_{\rm p}T_{\rm p}}{m_{\rm e}T_{\rm e}}} \exp \qty(-1/\beta_{\rm p}),
\end{equation}
where $c_1 = 0.92,~c_2 = 1.6 T_{\rm e}/T_{\rm p},$ and $c_3 = 18+5 \log_{10} (T_{\rm p}/T_{\rm e})$ for $T_{\rm p} > T_{\rm e}$, while $c_2 = 1.2 T_{\rm e}/T_{\rm p},$ and $c_3 = 18$ for $T_{\rm p} < T_{\rm e}$.
H10 has similar behaviour with K19 for $\beta_{\rm p} < 1$.
For K19, the proton-to-electron heating ratio saturate at $\sim 35$ at high-$\beta_{\rm p}$, while it monotonically increase with inverse $\beta_{\rm p}$ for H10.
The simulation setups are the same as the model B, except for numerical resolution and the coordinate.
We use the cylindrical coordinate ($r,\phi, z$), and the grid size is $\Delta_r = \Delta_z = 0.05$ kpc, which is half the size of our model B.

\begin{figure}
   \begin{center}
    \includegraphics[width=1\columnwidth]{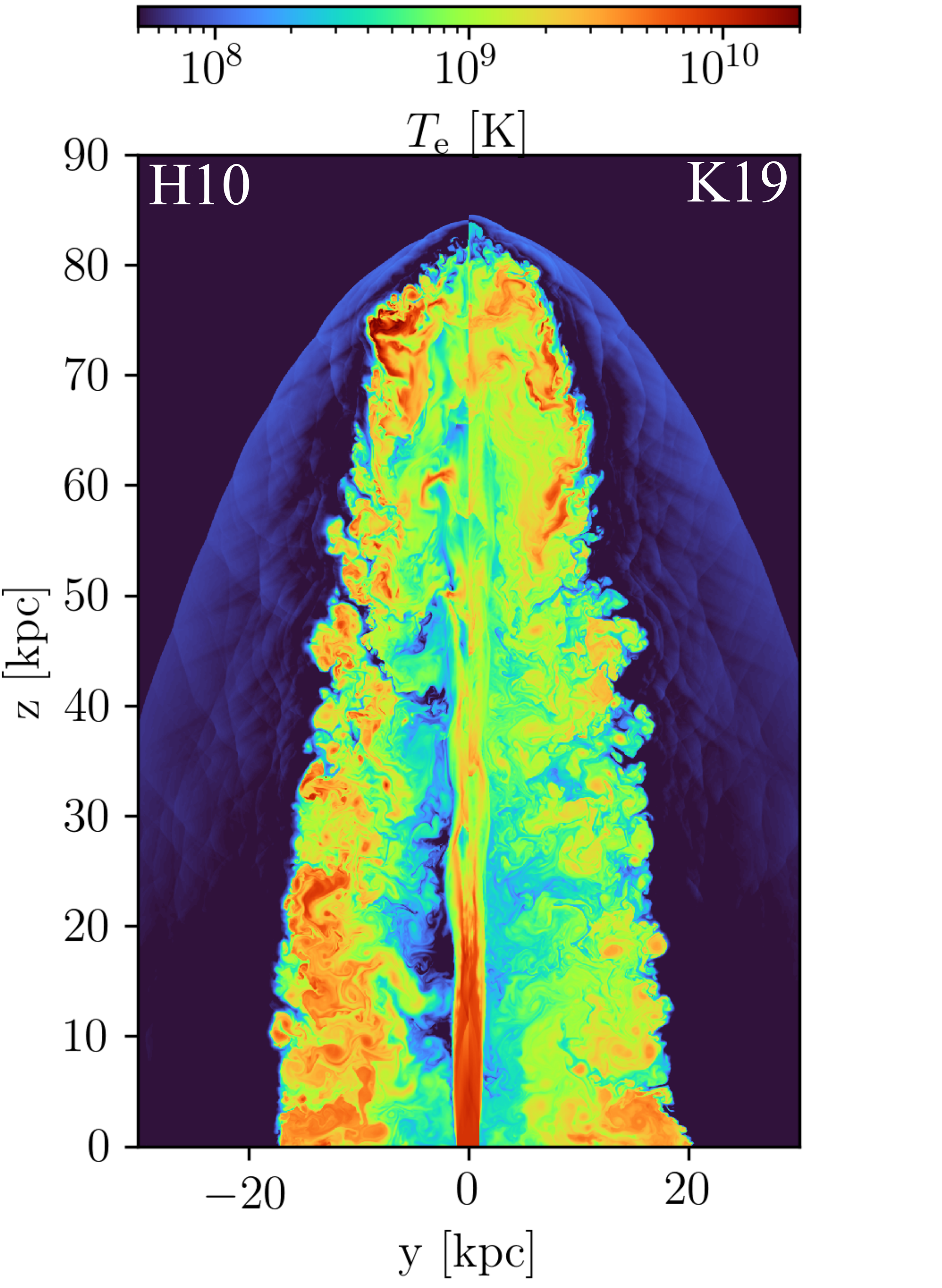}
     \caption{The distribution of the electron temperature using H10 (left side of the panel) and K19 (right side of the panel) at $t = 16.86$ Myr.} 
     \label{append_fig:1}
   \end{center}
 \end{figure}

\begin{figure}
   \begin{center}
    \includegraphics[width=1\columnwidth]{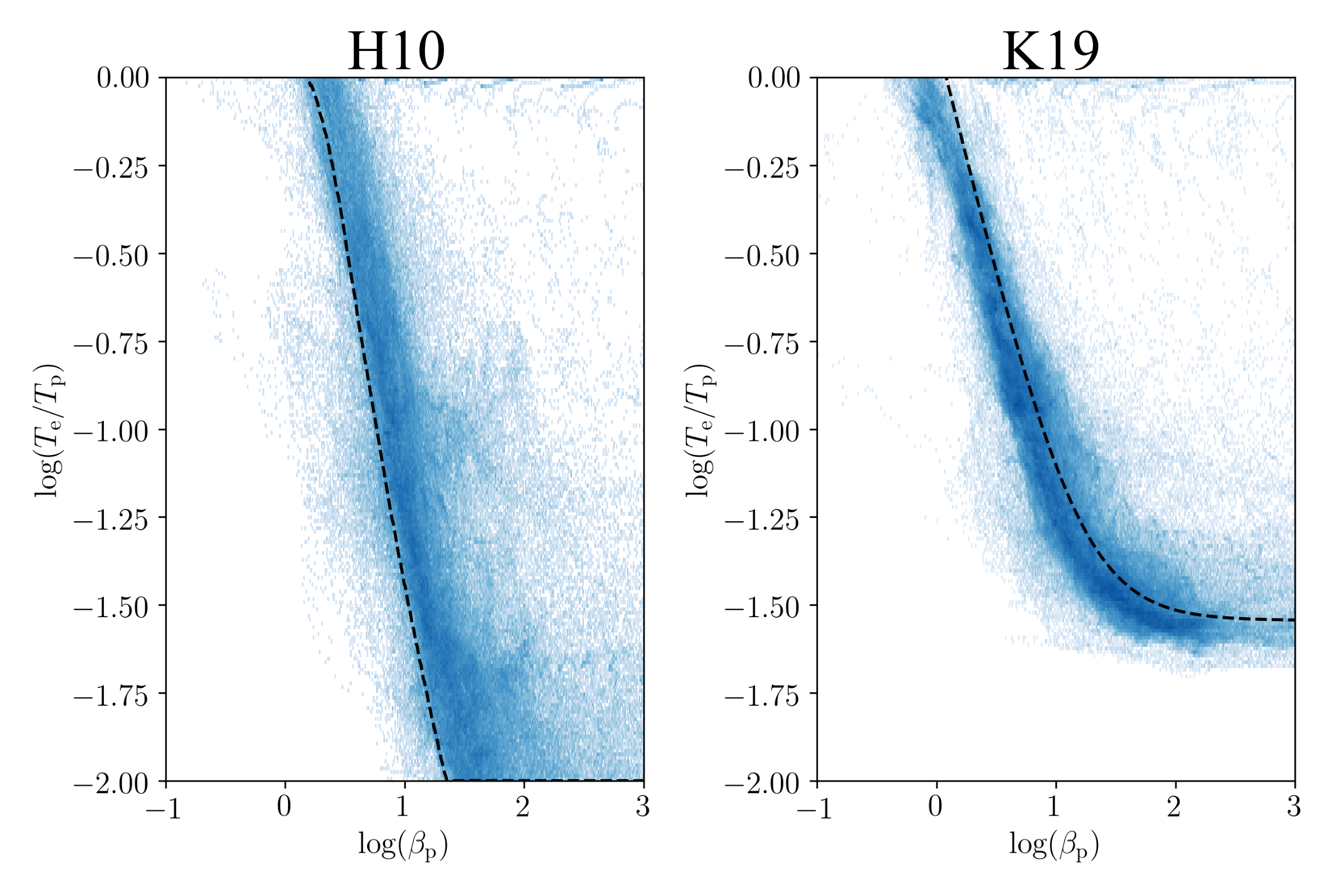}
     \caption{$T\rmf{e}/T\rmf{p} - \beta\rmf{p}$ histogram for regions in the cocoon using H10 (left) and K19 (right) at $t = 16.86$ Myr, respectively. The dashed line depicts the electron to proton temperature ratio corresponding to the equilibrium state for plasma $\beta\rmf{p}$, as implied by the turbulence heating in equation \ref{ch5_eq:turb_equi} for H10 and K19. } 
     \label{append_fig:2}
   \end{center}
 \end{figure}

In \figref{append_fig:1}, we show the distributions of the electron temperature using the different electron heating models (H10 and K19).
We find that the distributions of the electron temperature do not differ in the shocked-ICM. 
Meanwhile, the electron temperature of the model using H10 is slightly higher than that of the model using K19 in the cocoon, except for in the sheath of the beam.
Since the plasma-$\beta_{\rm p}$ in the sheath region is much higher than 10, the proton-to-electron heating ratio of H10 is significant low (see also \figref{append_fig:2}).
The heating ratio for both models is sensitive for plasma-$\beta_{\rm p}$, and hence the electron is locationally heated up around the jet head.
The ratio of the thermal electron energies, $U_{\rm e, H10}/U_{\rm e, K19}$, in the cocoon is 1.28 so that the choice of the sub-grid models would not have a significant affect for the main results in the case of the model A and B in Section \ref{sec:5}.
On the other hand, in the case of model C, the radio power using H10 would be much lower than that using K19 since the cocoon is consist of the high-beta plasma.
Comparison with the three-dimensional simulation of the model B, the propagation velocity in three-dimensional simulation is faster than that in axisymmetirc simulation due to the dentist-drill effect \citep{1974MNRAS.166..513S}.
This trend also found in previous three-dimensional simulation \citep{2019MNRAS.482.3718P}.
Because the kink mode do not develop in the axisymetric condition, the electrons are only heated up at the internal shocks. 
Meanwhile, in the three dimensional case, the dissipation due to the development of the kink mode also is dominant heating source for the electrons (see detail in section \ref{ch5_sec:dissipation}).

\section{Magnetic dissipative structures}
\label{sec:appendD}
\begin{figure*}
  \begin{center}
      \includegraphics[width=0.9\textwidth]{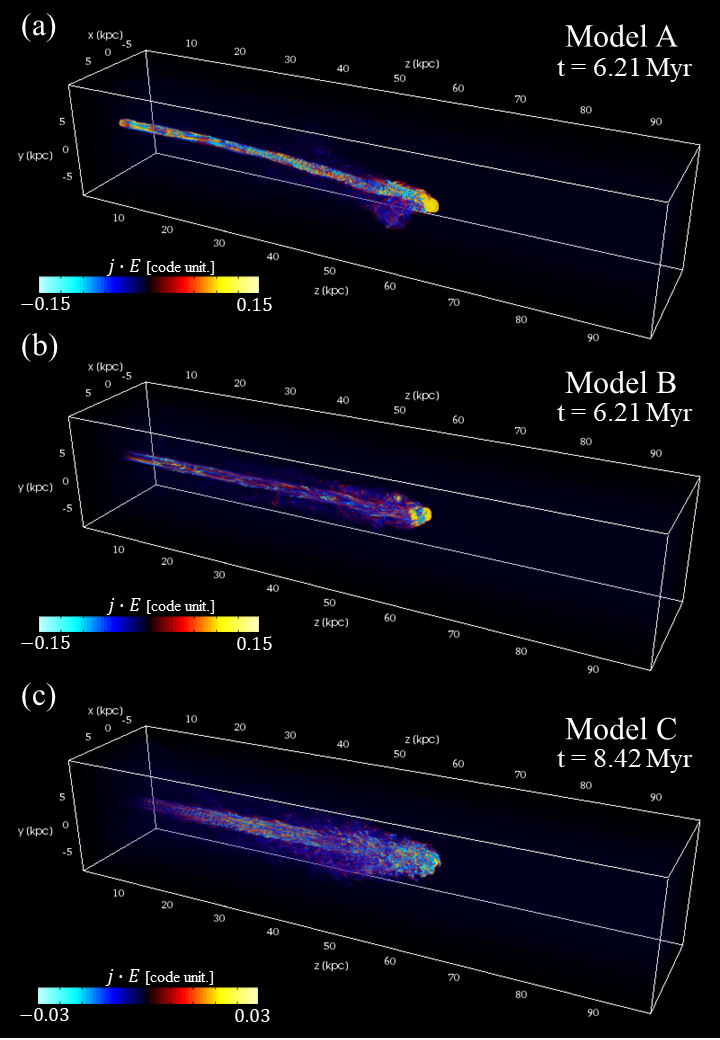}
    \caption{Three panels representing magnetic dissipative structures at early stages for all models. In each panel, we show  volume-renders with the strength of $\bm{J}\cdot \bm{E}$.}
  \label{append_fig:3}
  \end{center}
\end{figure*}

\begin{figure*}
  \begin{center}
       \includegraphics[width=0.9\textwidth]{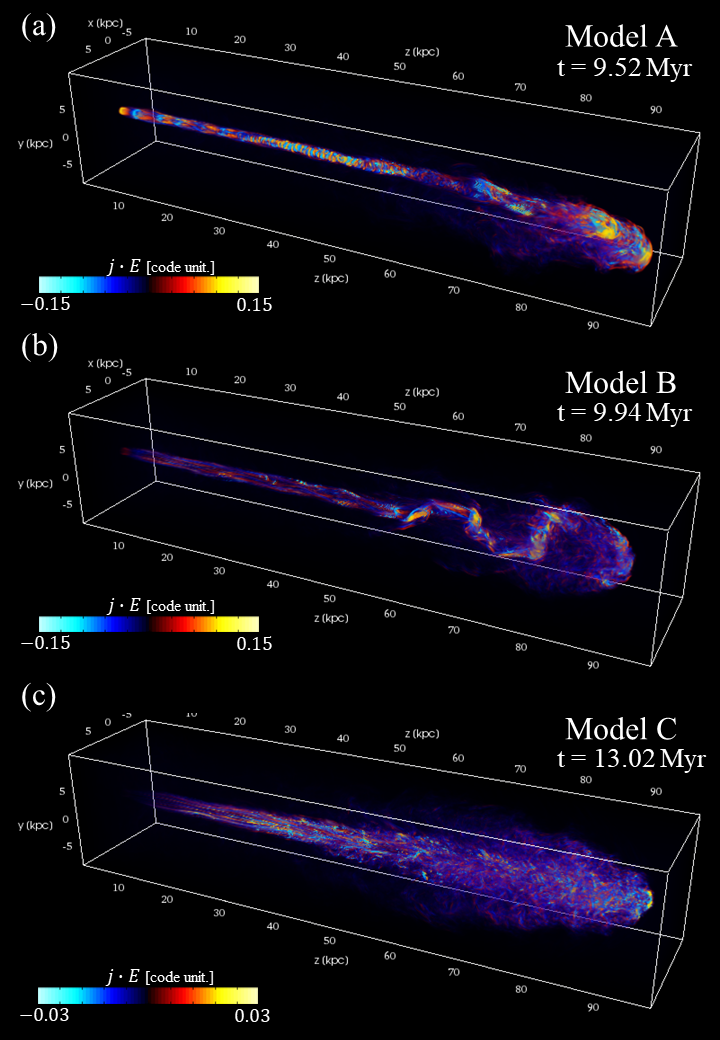}
    \caption{Same as \fullfigref{append_fig:3}, but at later phase.}
    \label{append_fig:4}
  \end{center}
\end{figure*}

\fullfigref{append_fig:3} and \figref{append_fig:4} shown magnetic dissipative structures for all models.

\end{appendix}
%
%

\end{document}